\documentclass[a4paper,11pt]{article}
\pdfoutput=1

\usepackage{jcappub} 

\usepackage[T1]{fontenc}
\usepackage{graphicx}
\usepackage{dcolumn}
\usepackage{bm}
\usepackage{amsmath}
\usepackage{amssymb}
\usepackage{slashed}
\usepackage{subcaption}
\usepackage{tikz} 
\usetikzlibrary{arrows.meta, positioning}

\title{Lyman-$\alpha$ Forest Constraint on Dark Matter from Dark Sector Decay}

\author[a]{Si-Yuan Zhao,}
\author[a]{Yi-Cheng Dai,}
\author[a]{Wei Liao,}
\author[b]{Yi-Song Lu}

\affiliation[a]{School of Physics, East China University of Science and Technology,\\ Shanghai 200237, China}
\affiliation[b]{School of Physics and Astronomy, Sun Yat-sen University (Zhuhai Campus),\\ Zhuhai 519082, China}

\emailAdd{siyuanzhao@mail.ecust.edu.cn}
\emailAdd{liaow@ecust.edu.cn}

\abstract{
By exploiting small-scale structure formation probed by Lyman-$\alpha$ forest observations, 
we study constraints on a model of dark matter from dark sector decay. 
We compute the phase space distribution of the dark matter and the linear matter power spectrum. 
We map the non-thermal dark matter distribution in this dark matter model to an approximate thermal warm dark matter distribution, 
and use this approximation to obtain a constraint from the Lyman-$\alpha$ forest observation.
We combine the latest Lyman-$\alpha$ forest bounds with the constraint from the Big Bang Nucleosynthesis. 
As these two probes offer highly complementary constraints, we impose strong limits on sub-GeV dark matter. 
Consequently, masses lighter than $\sim 10^{-1}\,\mathrm{GeV}$ are excluded, thereby significantly limiting the allowed parameter space.
More broadly, our findings demonstrate the utility of small-scale structure observations in testing non-thermal dark matter paradigms, offering valuable insights for exploring a wider class of late-time decay models.}


\begin{document}

\maketitle
\flushbottom

\section{Introduction}
\label{sec:intro}

Modern cosmological observations reveal that dark matter (DM) accounts for more than 80\% of the total matter content in the Universe~\cite{Planck2018}. Although weakly interacting massive particles (WIMPs) have long been the leading candidate for DM, decades of searches---ranging from collider experiments, e.g. at the Large Hadron Collider (LHC), to direct detection in underground laboratories and indirect detection via astronomical observations---have failed to detect any conclusive signal of DM~\cite{ATLAS:2021kxv, LZ:2022lsv, Fermi-LAT:2016uux}. Driven by this absence of signals, alternative models beyond the WIMP paradigm have garnered significant attention.

The absence of detection signals has raised serious concerns that the GeV-TeV scale WIMP DM should have interactions much weaker than previously thought~\cite{LZ:2022lsv, XENON:2023cxc, Meng:2021kio}, 
making an overabundance of the WIMP relic density in the standard thermal production mechanism hard to avoid.
One way out of this difficulty is to notice that the WIMP itself may not be a DM candidate, 
but rather the parent particle of the DM particle~\cite{Feng:2003xh, ChengYu:2021}. 
In such a scenario of light DM from the dark sector decay, the DM relic density is derived from the relic density of the dark sector WIMP.
Because the DM particle is significantly lighter than its parent dark sector WIMP, the final DM relic abundance is naturally suppressed compared to the standard thermal prediction.
Therefore, one can still have a weak scale interaction between the dark sector and the visible sector, 
and meanwhile obtain a DM relic density consistent with cosmological observations. 
A specific model of light DM from dark sector decay has been proposed~\cite{ChengYu:2021}, and  various aspects of this model have been explored, e.g. the constraints from the Big Bang Nucleosynthesis (BBN), Cosmic Microwave Background (CMB) and collider physics on this model~\cite{ChengYu:2021, Cheng_CPC:2022, Liu:2023_Shining, CMBSignature_paper, nuTHDM_paper}. 

Inherently, this model with a non-thermal production mechanism of the light DM is
capable of giving a large velocity dispersion of DM which could significantly suppress the
small-scale structure formation of the universe~\cite{Murgia:2017, Doroshkevich:1984}. 
The Lyman-$\alpha$ forest, which probes neutral hydrogen absorption lines in the Intergalactic Medium (IGM) within quasar spectra at redshifts $z \sim 2$\,--\,$6$, provides a high sensitive probe to the small-scale matter power spectrum in the quasi-linear regime and serves as a highly promising probe for such kind of DM~\cite{Viel:2005,Irsic2024}.
This work aims to study constraints on this model of light DM from dark sector decay~\cite{ChengYu:2021} 
utilizing the latest limits on small-scale structure suppression derived from Lyman-$\alpha$ forest observations.
We also include limits from other datasets. Our constraint method is generic and applies to all of them.

The paper is organized as follows: In section~\ref{sec:model}, we describe in detail the model of light DM from the dark sector decay. 
We study the decay kinematics and the momentum distribution of the DM in this model 
using the Boltzmann equations that govern the phase space evolution of DM distribution in the early universe. 
In section~\ref{sec:imprint}, we study the thermal Warm Dark Matter (WDM) approximation,
i.e.,  approximating the DM distribution, which is basically non-thermal, as a thermal Warm Dark Matter (WDM) distribution.
In section~\ref{sec:results}, we formalize the methodologies to constrain this DM model and present numerical results
of constraints from the BBN and the Lyman-$\alpha$ forest observations on this DM model.
Finally, our conclusions are summarized in section~\ref{sec:conclusions}.

\section{Momentum Distribution of DM from Dark Sector Decay}
\label{sec:model}
The model under consideration~\cite{ChengYu:2021} includes a scalar mediator $\phi$, alongside a DM particle $\chi$ and a right-handed neutrino $N_R$. 
$\chi$  is a Dirac fermion which does not couple directly to the Standard Model (SM) particles.
$N_R$ is a fermion with a Majorana mass, which couples to the SM particles through the Yukawa couplings with the SM Higgs doublet.
The DM  $\chi$ is stabilized by an imposed $Z_2$ symmetry. The total Lagrangian is given by
\begin{equation}
    \mathcal{L} = \mathcal{L}_{\mathrm{SM}} + \mathcal{L}_{\mathrm{Seesaw}} + \mathcal{L}_{\mathrm{DS}} + \mathcal{L}_{\mathrm{int}} \,,
\end{equation}
with new physics terms defined as
\begin{align}
    \mathcal{L}_{\mathrm{Seesaw}} &= -Y_{\alpha} \bar{L}_\alpha \tilde{H} N_{R} - \frac{1}{2} M_N \bar{N}_R^c N_R + \mathrm{h.c.} \,, \\
    \mathcal{L}_{\mathrm{DS}} &= \bar{\chi}(i\slashed{\partial} - m_\chi)\chi + \frac{1}{2}\partial_\mu\phi\partial^\mu\phi - \frac{1}{2}\mu_\phi^2\phi^2 - \frac{1}{4}\lambda_\phi\phi^4 - \lambda_{H\phi}H^\dagger H\phi^2 \,, \\
    \mathcal{L}_{\mathrm{int}} &= -y_{\mathrm{DS}}\phi\bar{\chi}N_R + \mathrm{h.c.} \,,
\end{align}
where $L_\alpha$ and $H$ (with $\tilde{H} = i\tau_2 H^*$) denote the SM left-handed lepton and the Higgs doublet, respectively. 
For simplicity, we consider the scenario with a single right-handed neutrino $N_R$. 
For three right-handed neutrinos with degenerate Majorana mass, the analysis would be similar to what follows.
The DM $\chi$ couples to other particles of the dark sector through coupling $y_{\mathrm{DS}}$.
We assume that $y_{\mathrm{DS}}$ is very small and $\chi$ is never in thermal equilibrium with the thermal bath.

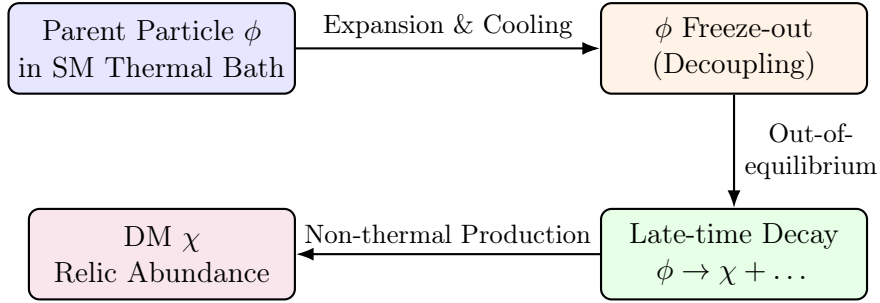
\begin{figure}[htbp]
    \centering
    \begin{tikzpicture}[
        node distance=1.5cm and 4cm, 
        box/.style={draw, rectangle, rounded corners, minimum width=3.5cm, minimum height=1.2cm, align=center, thick},
        arrow/.style={-Latex, thick}
    ]
        \node[box, fill=blue!10] (thermal) {Parent Particle $\phi$ \\ in SM Thermal Bath};
        \node[box, fill=orange!10, right=of thermal] (freezeout) {$\phi$ Freeze-out \\ (Decoupling)};
        
        \node[box, fill=green!10, below=of freezeout] (decay) {Late-time Decay \\ $\phi \to \chi + \dots$};
        \node[box, fill=purple!10, left=of decay] (relic) {DM $\chi$ \\ Relic Abundance};

        \draw[arrow] (thermal) -- node[above, font=\small] {Expansion \& Cooling} (freezeout);
        \draw[arrow] (freezeout) -- node[right, font=\small, align=center] {Out-of-\\equilibrium} (decay);
        \draw[arrow] (decay) -- node[above, font=\small] {Non-thermal Production} (relic);
    \end{tikzpicture}
    \caption{Schematic flowchart of the $\chi$ DM production mechanism. The heavy parent particle $\phi$ first freezes out from the SM thermal bath and subsequently decays into the DM particle $\chi$.}
    \label{fig:ds_flowchart}
\end{figure}

The dark scalar $\phi$ couples to the SM directly via a weak-scale coupling constant $\lambda_{H\phi}$.
After electroweak symmetry breaking, the scalar $\phi$ acquires an additional mass contribution
\begin{equation}
    m_\phi^2 = \mu_\phi^2 + \lambda_{H\phi}v_{\mathrm{ew}}^2 \,,
\end{equation}
where $v_{\mathrm{ew}}/\sqrt{2}$ is the vacuum expectation value of the Higgs doublet.
Assuming $\phi$ is much heavier than both $\chi$ and $N_R$, the decay channel $\phi \to \chi N_R$ is kinematically allowed. 
Driven by the $\lambda_{H\phi}H^\dagger H\phi^2$ interaction, $\phi$ can establish thermal equilibrium with the SM thermal bath in the early Universe. As the temperature drops below the mass scale of $\phi$, it freezes out and decouples from the thermal bath. 
Its subsequent decay dominates the production mechanism of DM and ultimately yields the final relic abundance of $\chi$.
Fig. \ref{fig:ds_flowchart} gives a schematic flowchart of this DM production mechanism.

$\chi$ DM can also be produced in the early universe through the freeze-in mechanism, i.e. through the annihilation of $\phi$ and $N_R$ in the thermal bath.
It has been shown that as long as $y_{\mathrm{DS}} \lesssim 2 \times 10^{-12}$, the contribution of freeze-in mechanism to the relic density of $\chi$ 
is bounded to $\lesssim 10\%$~\cite{ChengYu:2021}.
Moreover, freeze-in production of $\chi$ occurs at extremely high temperatures within the early thermal bath. 
Consequently, these $\chi$ particles from the freeze-in production undergo significantly more cosmic redshift 
compared to the $\chi$ particles produced via late-time $\phi$ decays. 
This implies that the freeze-in component is substantially ``colder'' than the decay component.
Therefore, even near the parameter space boundary ($y_{\mathrm{DS}} \lesssim 2 \times 10^{-12}$) where the freeze-in contribution might marginally reach $\mathcal{O}(10\%)$, the suppression of the small-scale power spectrum and the free-streaming length are overwhelmingly dictated by the dominant and ``hotter'' decay component. Motivated by this picture, we will neglect the contribution of the freeze-in mechanism and concentrate on the decay production mechanism.

The phenomenology of this model is primarily governed by five free parameters: the parent particle mass $m_\phi$, the DM mass $m_\chi$, the right-handed neutrino mass $m_N$, the decay coupling constant $y_{\mathrm{DS}}$ and the thermal coupling $\lambda_{H\phi}$. 
$\lambda_{H\phi}$ determines the abundance of $\phi$ after freeze-out. Assuming that the $\chi$ particles constitute the entirety of the observed DM ($\Omega_\chi = \Omega_{\mathrm{DM}}$), the mechanism of light DM from dark sector decay dictates
\begin{equation}
    \Omega_\chi = \frac{m_\chi}{m_\phi} \Omega_{\phi,\mathrm{FO}}(m_\phi, \lambda_{H\phi}) \,,
    \label{eq:Omega}
\end{equation}
where $\Omega_{\phi,\mathrm{FO}}$ denotes the hypothetical present-day relic abundance of $\phi$ assuming they were stable. Since the late-time decay ($\phi \rightarrow N + \chi$) occurs well after freeze-out, the comoving number of particles translates one-to-one from $\phi$ to $\chi$. Therefore, the final DM density is simply $\Omega_{\phi,\mathrm{FO}}$ rescaled by the mass ratio $m_\chi / m_\phi$.
Consequently, by fixing the theoretical DM relic abundance to match the current observational value, the thermal coupling $\lambda_{H\phi}$ is determined. Our subsequent phenomenological analysis focuses primarily on constraining $m_\phi$, $m_\chi$, and $y_{\mathrm{DS}}$, while treating the right-handed neutrino mass $m_N$ as a free parameter.

To evaluate the non-thermal phase space distribution of the DM $\chi$, we track the evolution of the dark sector particles in the late-time decay phase. The Boltzmann equations include the Hubble expansion term $H \equiv \dot{a}/a$ to account for the physical momentum redshift:
\begin{align}
    \frac{\partial f_\phi}{\partial t} - H p_\phi \frac{\partial f_\phi}{\partial p_\phi} &= C_\phi \,, \label{eq:boltz_full_phi} \\
    \frac{\partial f_\chi}{\partial t} - H p_\chi \frac{\partial f_\chi}{\partial p_\chi} &= C_\chi \,, \label{eq:boltz_full_chi}
\end{align}
where $f_\phi(p_\phi, t)$ and $f_\chi(p_\chi, t)$ are the phase space distribution functions of the $\phi$ and $\chi$ particles, respectively. The interaction terms $C_\phi$ and $C_\chi$ include all relevant interactions:
\begin{align}
    C_\phi &= C_{\phi\,\mathrm{SM} \to \phi\,\mathrm{SM}} - C_{\phi\phi\to \mathrm{SM}} - C_{\phi\phi\to NN} - C_{\phi\to N\chi} \,, \\
    C_\chi &= C_{\phi\phi\to\chi\chi} + C_{NN\to\chi\chi} + C_{hN\to\phi\chi} + C_{h\nu\to\phi\chi} + C_{\phi\to N\chi} \,.
\end{align}
The collision term $C_{\phi\phi\to \mathrm{SM}}$ describes the annihilation of $\phi$ into SM particles, which can be safely neglected after $\phi$ decouples from the thermal bath. $C_{\phi\phi\to NN}$, $C_{\phi\phi\to\chi\chi}$, $C_{NN\to\chi\chi}$, $C_{hN\to\phi\chi}$, and $C_{h\nu\to\phi\chi}$, are proportional to higher orders of the decay coupling (e.g., $y_{\mathrm{DS}}^4$), and can all be neglected for $y_{\mathrm{DS}} \lesssim 10^{-12}$~\cite{ChengYu:2021, FISW:2021}. 

As with typical WIMPs, the elastic scattering term $C_{\phi\,\mathrm{SM} \to \phi\,\mathrm{SM}}$ keeps $\phi$ in kinetic equilibrium with the thermal bath after its freeze-out. If this scattering rate is insufficient, 
one can solve the Boltzmann equation for $f_\phi$ and $f_\chi$ simultaneously. If this process is efficient, we can simply assume that $f_\phi$ maintains a thermal momentum profile tracking the bath temperature, with its overall normalization determined by the evolving number density which can be solved analytically. A detailed discussion is provided in Appendix~\ref{app:calculation.2} and~\ref{app:approximation.1}.
For simplicity, we take the later scenario in the following discussion.
Consequently, at late times, the Boltzmann equation for $\chi$ is dominated by the decay term $C_{\phi\to N\chi}$, yielding
\begin{align}
    \frac{\partial f_\chi}{\partial t} - H p_\chi \frac{\partial f_\chi}{\partial p_\chi} &= \frac{m_\phi^2 \Gamma_\phi}{2p^* E_\chi(p_\chi)} \int_{p_{\phi-}}^{p_{\phi+}} dp_\phi \frac{f_\phi(p_\phi) p_\phi}{E_\phi(p_\phi)} \,. \label{eq:boltz_simp_chi}
\end{align}
Here, $p^*$ denotes the magnitude of momentum of the DM particle evaluated in the rest frame of parent $\phi$ particle. 
$\Gamma_\phi$  is the decay rate of $\phi$ in the rest-frame, as detailed in Appendix~\ref{app:calculation.1}.
 The integration limits $p_{\phi\pm}$ represent the maximum and minimum momenta of $\phi$ in the cosmic frame that can kinematically yield a DM $\chi$ with momentum $p_\chi$.

To absorb the cosmic expansion terms on the left-hand side of Eq.~(\ref{eq:boltz_simp_chi}), we transform to the comoving momentum $q \equiv a p$ :
\begin{equation}
    F(q) \equiv g q^3 f(p) / (2\pi^2),
\end{equation}
which gives the comoving number density as
\begin{equation}
    N \equiv a^3 n = \int d\ln q \, F(q)\,.
\end{equation}
Under this transformation, the partial time derivative at constant $q$ naturally eliminates the Hubble term, significantly simplifying the numerical evaluation. 
The Boltzmann equation finally reduce to
\begin{align}
    \frac{\partial F_\chi(q_\chi)}{\partial t} &= \frac{m_\phi^2 \Gamma_\phi q_\chi^2}{2 a p^* E_\chi(q_\chi)} \int d\ln q_\phi \frac{F_\phi(q_\phi)}{ q_\phi E_\phi(q_\phi)} \,. \label{eq:Cchi_F}
\end{align}

To further optimize computational efficiency, we adopt the zero-momentum approximation. 
The validity of this approximation renders the exact kinetic equilibrium status of $\phi$ irrelevant to the final results.
As detailed in Appendix~\ref{app:approximation.1}, this approximation is highly accurate. By numerically solving the resulting equation, we obtain the non-thermal phase space distribution function of $\chi$.

\begin{figure}[htbp]
    \centering
    \begin{subfigure}[b]{0.32\textwidth}
        \centering
        \includegraphics[height=4cm]{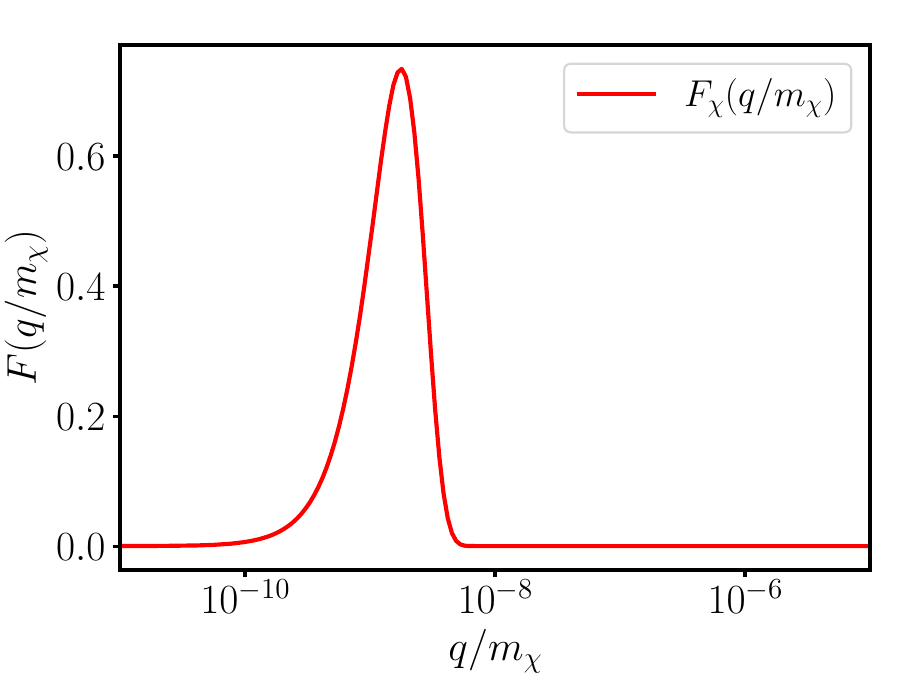} 
        \caption{$m_\chi=10\,\mathrm{GeV}$}
    \end{subfigure}
    \hfill 
    \begin{subfigure}[b]{0.30\textwidth}
        \centering
        \includegraphics[height=4cm]{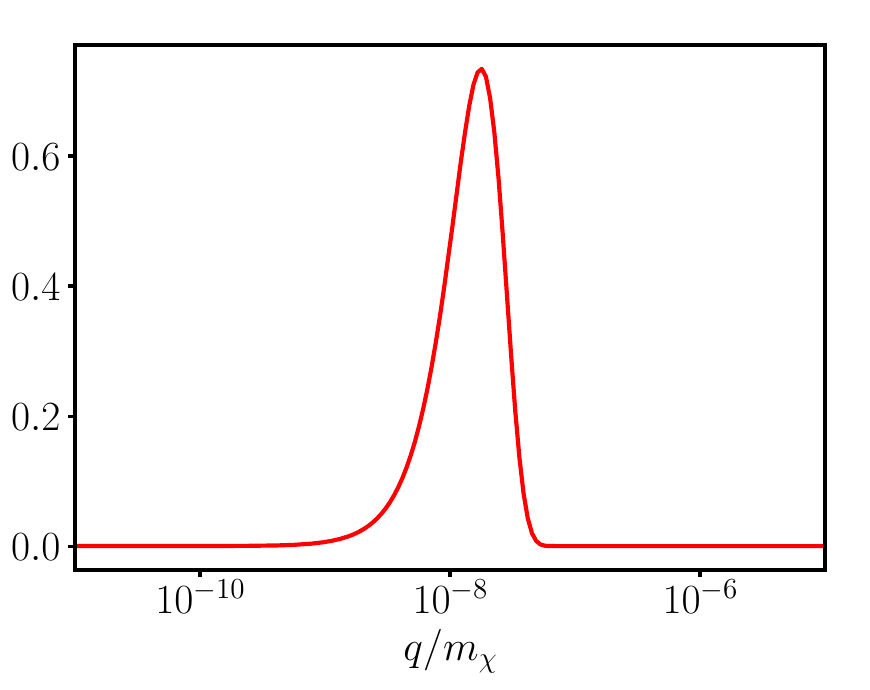} 
        \caption{$m_\chi=1\,\mathrm{GeV}$}
    \end{subfigure}
    \hfill    
     \begin{subfigure}[b]{0.32\textwidth}
        \centering
        \includegraphics[height=4cm]{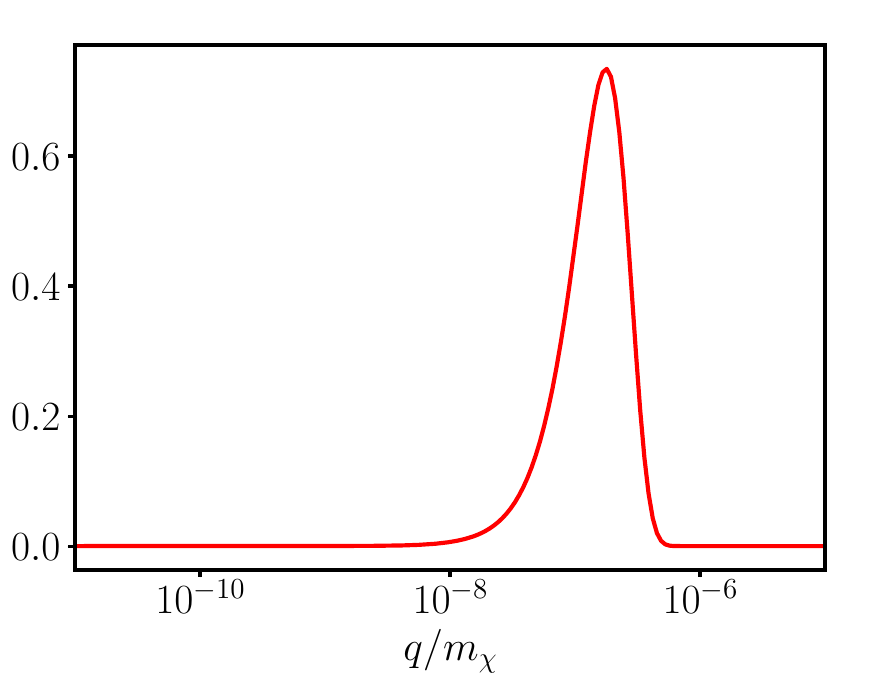} 
        \caption{$m_\chi=0.1\,\mathrm{GeV}$}
    \end{subfigure}
    \hfill
    \begin{subfigure}[b]{0.32\textwidth}
        \centering
        \includegraphics[height=4cm]{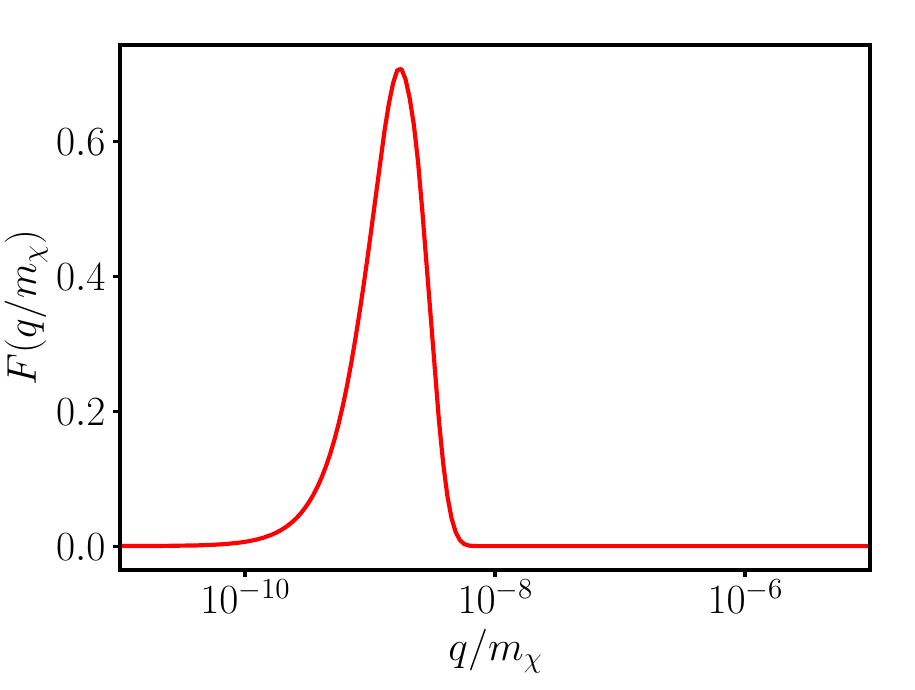} 
        \caption{$m_\phi=10\,\mathrm{GeV}$}
    \end{subfigure}
    \hfill 
    \begin{subfigure}[b]{0.30\textwidth}
        \centering
        \includegraphics[height=4cm]{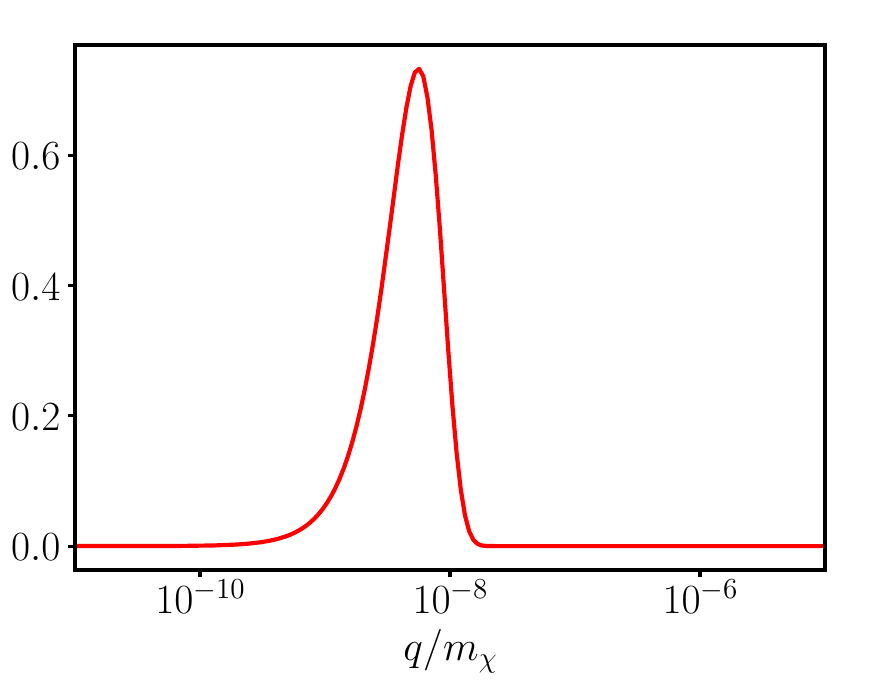} 
        \caption{$m_\phi=100\,\mathrm{GeV}$}
    \end{subfigure}
    \hfill    
     \begin{subfigure}[b]{0.32\textwidth}
        \centering
        \includegraphics[height=4cm]{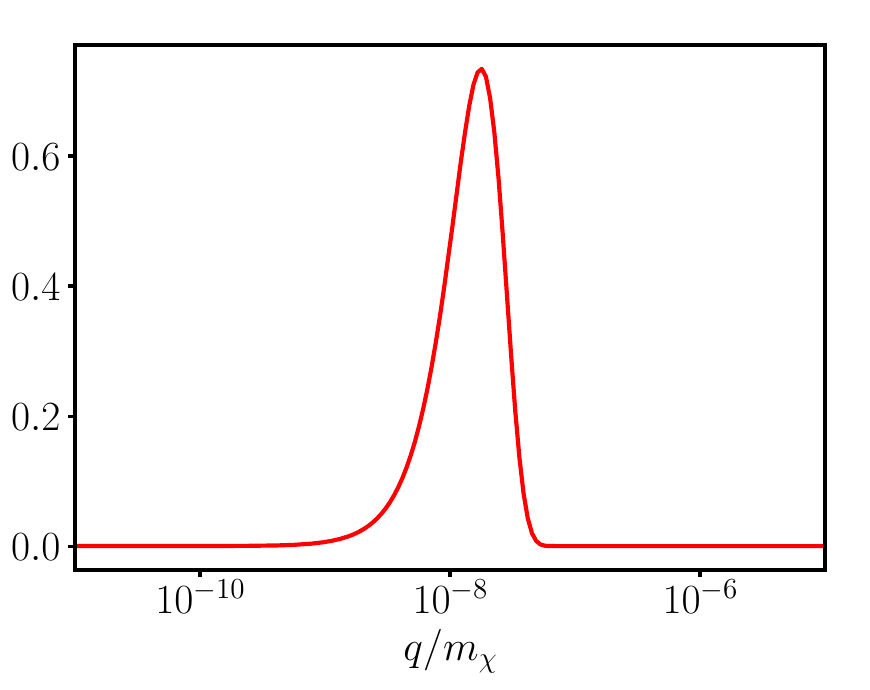} 
        \caption{$m_\phi=1000\,\mathrm{GeV}$}
    \end{subfigure}

    \caption{Unperturbed background comoving phase-space distribution of the $\chi$ DM after production is completed. The horizontal axis represents the dimensionless comoving momentum $q/m_\chi$. In all panels, the parameters are fixed at $y_{\mathrm{DS}} = 10^{-12}$ and $m_N = 1\,\mathrm{GeV}$. The upper panels illustrate the distributions for a fixed $m_\phi = 1000\,\mathrm{GeV}$ with varying $m_\chi \in \{10, 1, 0.1\}\,\mathrm{GeV}$. The lower panels display the scenarios for a fixed $m_\chi = 1\,\mathrm{GeV}$ with varying $m_\phi \in \{10, 100, 1000\}\,\mathrm{GeV}$.}
\label{fig:Fchi1} 
\end{figure}

\begin{figure}[htbp]
    \centering
    \begin{subfigure}[b]{0.32\textwidth}
        \centering
        \includegraphics[height=4cm]{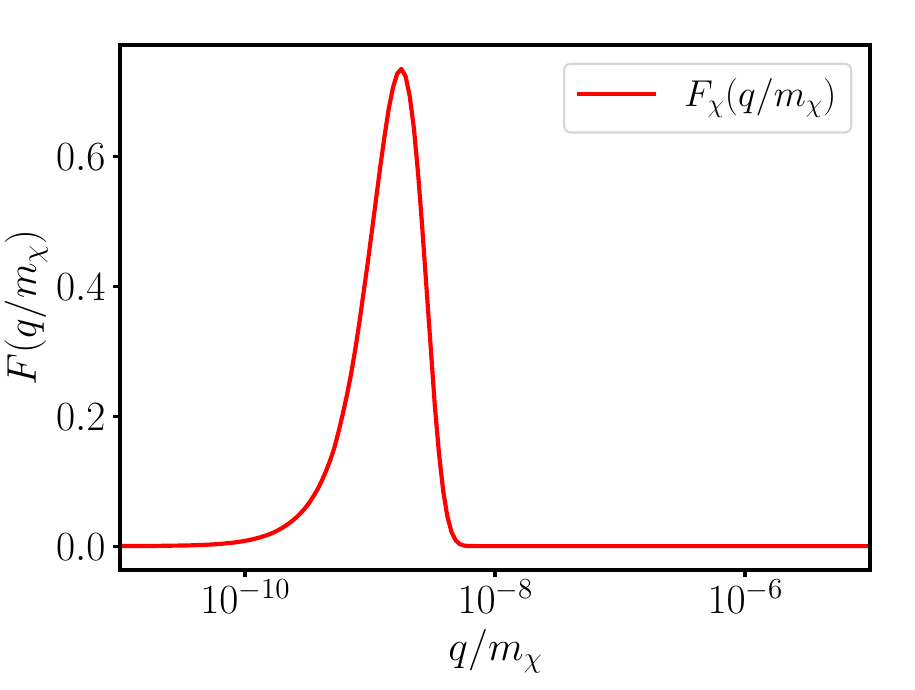} 
        \caption{$y_{\mathrm{DS}} = 10^{-11}$}
    \end{subfigure}
    \hfill 
    \begin{subfigure}[b]{0.30\textwidth}
        \centering
        \includegraphics[height=4cm]{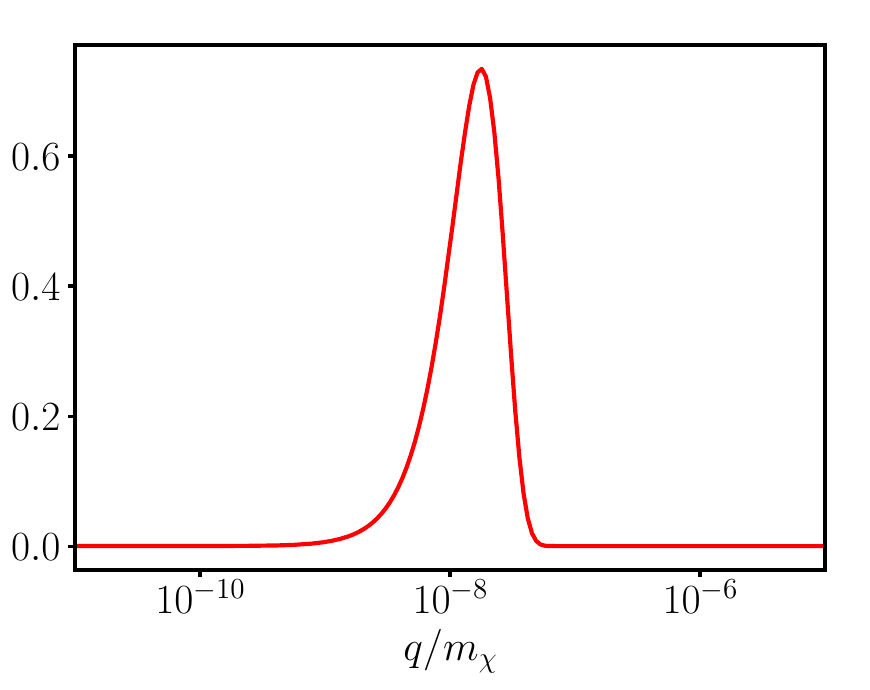} 
        \caption{$y_{\mathrm{DS}} = 10^{-12}$}
    \end{subfigure}
    \hfill    
     \begin{subfigure}[b]{0.32\textwidth}
        \centering
        \includegraphics[height=4cm]{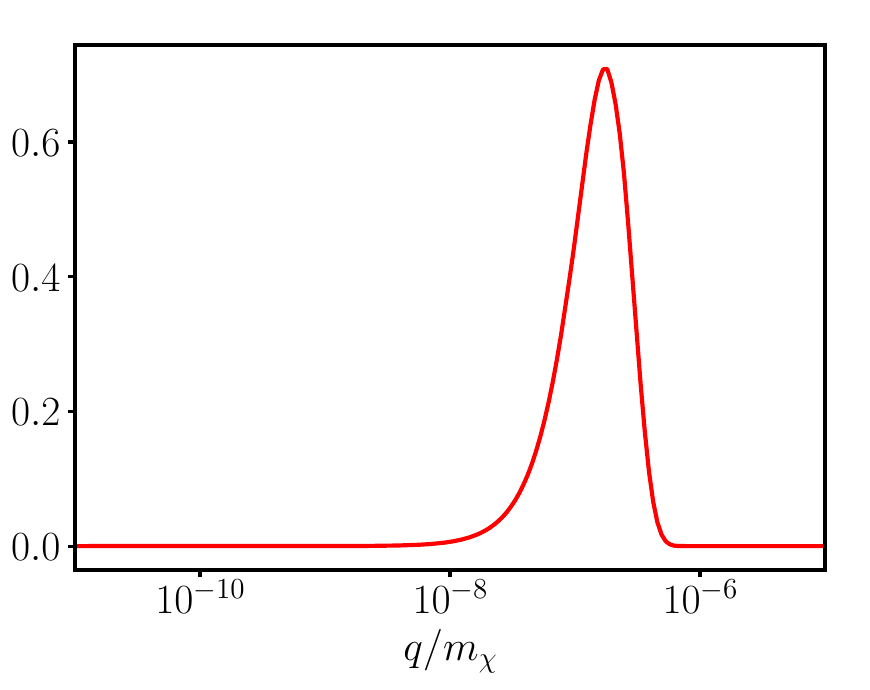} 
        \caption{$y_{\mathrm{DS}} = 10^{-13}$}
    \end{subfigure}

    \caption{Same as Fig.~\ref{fig:Fchi1}, but for varying decay couplings $y_{\mathrm{DS}} \in \{10^{-11}, 10^{-12}, 10^{-13}\}$. The other parameters are fixed at $m_\phi = 1000\,\mathrm{GeV}$, $m_\chi = 1\,\mathrm{GeV}$, and $m_N = 1\,\mathrm{GeV}$.}
    \label{fig:Fchi2}
\end{figure}

Fig.~\ref{fig:Fchi1} and Fig.~\ref{fig:Fchi2} present the unperturbed background phase-space distributions of the DM versus the dimensionless comoving momentum $q/m_\chi$.
Here, the term ``unperturbed background'' signifies that the momentum distributions are evaluated in a homogeneous and isotropic universe. They represent the global cosmological averages, explicitly excluding any local density fluctuations or peculiar velocity flows induced by non-linear gravitational clustering.
We use the comoving momentum $q \equiv ap$ (with the scale factor normalized to $a=1$ at the present epoch) because it is conserved for freely propagating particles in the homogeneous background once the decay concludes, thereby freezing the primordial kinematic shape that sources linear structure formation, regardless of late-time non-linear clustering.

As depicted in Fig.~\ref{fig:Fchi1}, the distribution of the dimensionless comoving momentum $q/m_\chi$ is highly sensitive to the mass hierarchy between the parent particle $\phi$ and the DM $\chi$. The upper panels demonstrate that decreasing $m_\chi$ from $10\,\mathrm{GeV}$ to $0.1\,\mathrm{GeV}$ significantly shifts the distribution peak towards higher values (from $q/m_\chi \sim 10^{-9}$ to $\sim 10^{-7}$). 
Physically this is because a larger mass gap produces DM particles with much higher initial momentum.
Similarly, the lower panels show that increasing the parent mass $m_\phi$ from $10\,\mathrm{GeV}$ to $1000\,\mathrm{GeV}$ also drives the peak to higher $q/m_\chi$. 

Fig.~\ref{fig:Fchi2} shows how the DM phase space distribution depends on the coupling $y_{\mathrm{DS}}$. A larger $y_{\mathrm{DS}}$ increases the decay rate $\Gamma_\phi$, so that $\phi$ decay and DM production occurs at earlier times.
Since the scale factor $a$ at earlier decay is then smaller, the resulting DM particles have smaller comoving momenta ($q \propto p_{\mathrm{initial}} \cdot a_{\mathrm{decay}}$). This effect shifts the distribution peak to lower values, effectively yielding a colder DM relic at later times.

\section{Thermal WDM approximation to DM distribution}
\label{sec:imprint}

After being produced from late-time decays, the DM particles possess substantial momenta. If their velocities remain sufficiently large at late times, the free-streaming effect would hinder small-scale clustering. The imprints left by the DM on structure formation can be characterized in two main ways~\cite{FISW:2021, FI:2025}: (1) the velocity dispersion $\langle v^2 \rangle$, and (2) the transfer function $T(k)$. Both can be constrained using existing limits derived from Lyman-$\alpha$ forest data.

Rather than relying on $N$-body simulations, we constrain our parameter space by mapping the model to established thermal WDM limits. Although the $\chi$ DM particles possess a non-thermal distribution resulting from late-time decay, their continuous distribution features a single dominant peak, as shown in Fig.~\ref{fig:Fchi1} and Fig.~\ref{fig:Fchi2} in Sec.~\ref{sec:model}. This single-peaked structure makes its impact on small-scale structure formation governed primarily by its velocity dispersion~\cite{FISW:2021,Rev:2021}, rendering it highly analogous to that of thermal WDM.

Methodologically, our work builds upon the equivalent mapping approach introduced in Ref.~\cite{FISW:2021}. 
However, while their study relies on a strong mass-hierarchy approximation to derive an analytical expression for the DM momentum distribution, our work relaxes this assumption by directly evaluating the numerical phase-space distribution.
This extension allows us to safely explore the highly mass-degenerate limits $m_N/m_\phi \to 1$ and $m_\chi/m_\phi \to 1$. In this degenerate regime, extreme phase-space suppression significantly prolongs the decay lifetime. We reveal that the prolonged lifetime in this degenerate regime is strictly forbidden by BBN.

With this numerical framework, the parameter space of our model can be robustly mapped to an equivalent thermal WDM model, allowing us to directly translate the Lyman-$\alpha$ constraints. In this section, we study this equivalent WDM approximation to our DM model in two aspects, i.e. the velocity dispersion and the linear transfer function.
 
\subsection{Velocity Dispersion}
\label{subsec:velocity_wdm}

In the hydrodynamic description, the velocity dispersion of the DM fluid acts as an effective pressure that resists gravitational collapse. This resistance scale is characterized by the Jeans length $\lambda_J$, defined as~\cite{Dodelson:2020}
\begin{equation}
    \lambda_J \propto c_s \sqrt{\frac{\pi}{G\rho}} \,,
\end{equation}
where $G$ is the Newton's gravitational constant, $\rho$ is the background matter density, and $c_s$ is the sound speed of the fluid. The square of the sound speed is given by the ratio of kinetic pressure $P$ to density $\rho$ ($c_s^2 \sim P/\rho$), which directly scales with the mean squared velocity of the DM particles ($c_s^2 \propto \langle v^2 \rangle$). 

To constrain the DM model, we adopt the equivalent thermal WDM approximation~\cite{FISW:2021}. The impact of our non-thermal DM on structure formation can be evaluated by mapping its unperturbed background kinematics to those of a thermal relic WDM. This equivalent thermal WDM shares the same relic energy density and pseudo-velocity dispersion. The effective parameters of this WDM model, denoted by the subscript $\mathrm{eff}$, are determined by the following matching relations
\begin{align}
    m_{\mathrm{\chi}} \cdot n_{\mathrm{\chi}} &= m_{\mathrm{eff}} \cdot n_{\mathrm{eff}}(T_{\mathrm{eff}}) \,, \label{eq:abundance} \\
    \frac{\langle q^2 \rangle_{\mathrm{\chi}}}{m_{\mathrm{\chi}}^2} &= \frac{\langle q^2 \rangle_{\mathrm{eff}}(T_{\mathrm{eff}})}{m_{\mathrm{eff}}^2} \,, \label{eq:momentum}
\end{align}
Here, $n$ and $\langle q^2 \rangle$ denote the unperturbed background number density and the mean squared comoving momentum, respectively. For the equivalent thermal WDM, both $n_{\mathrm{eff}}$ and $\langle q^2 \rangle_{\mathrm{eff}}$ are derived from a Fermi-Dirac distribution and depend solely on the effective temperature $T_{\mathrm{eff}}$. Solving these two equations yields both the effective temperature $T_{\mathrm{eff}}$ and the effective thermal mass $m_{\mathrm{eff}}$, the latter of which can then be directly compared against the established Lyman-$\alpha$ constraints~\cite{Irsic2024}.
Because the comoving momentum $q$ is conserved after DM production, matching the comoving pseudo-velocity dispersion $\langle q^2 \rangle / m^2$ ensures that the physical non-relativistic velocity dispersions at later time can be obtained directly using
$\langle v^2 \rangle \simeq \langle q^2 \rangle / (a^2 m^2)$. 

\begin{figure}[htbp]
    \centering
        \begin{subfigure}[b]{0.32\textwidth}
        \centering
        \includegraphics[height=4cm]{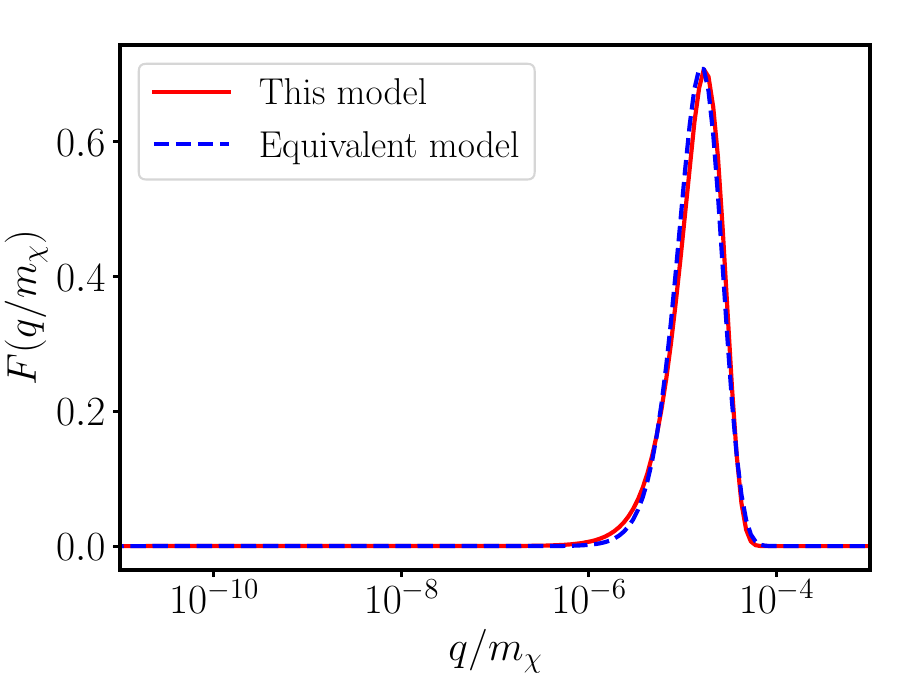} 
        \caption{$m_{\chi}=0.01\, \mathrm{GeV}$}
    \end{subfigure}
    \hfill 
    \begin{subfigure}[b]{0.30\textwidth}
        \centering
        \includegraphics[height=4cm]{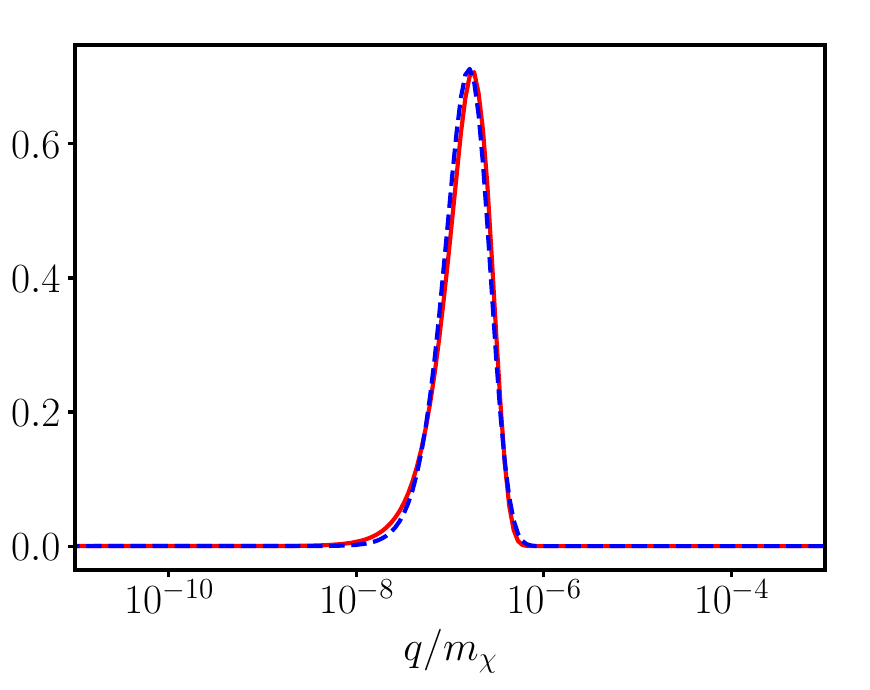} 
        \caption{$m_{\chi}=1\,\mathrm{GeV}$}
    \end{subfigure}
    \hfill 
    \begin{subfigure}[b]{0.32\textwidth}
        \centering
        \includegraphics[height=4cm]{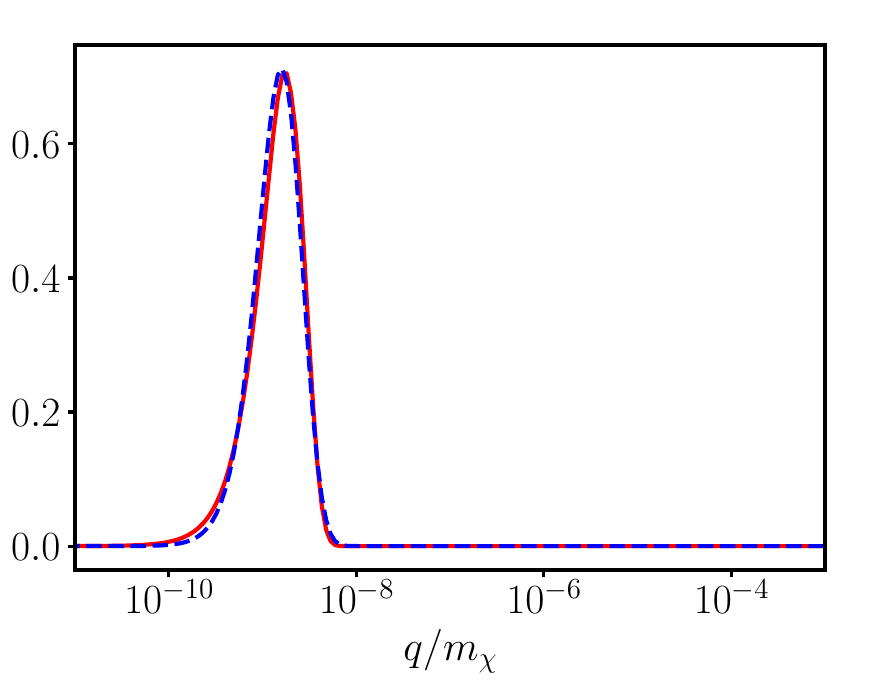} 
        \caption{$m_{\chi}=100\,\mathrm{GeV}$}
    \end{subfigure}
    \hfill 
    \begin{subfigure}[b]{0.32\textwidth}
        \centering
        \includegraphics[height=4cm]{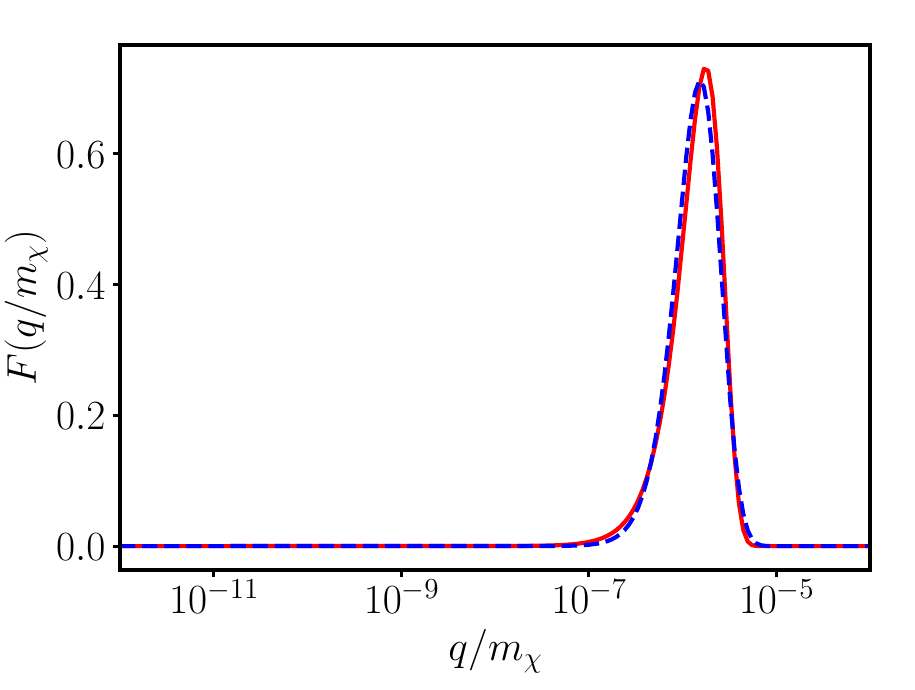} 
        \caption{$m_{\chi}=0.01\, \mathrm{GeV}$}
    \end{subfigure}
    \hfill 
    \begin{subfigure}[b]{0.30\textwidth}
        \centering
        \includegraphics[height=4cm]{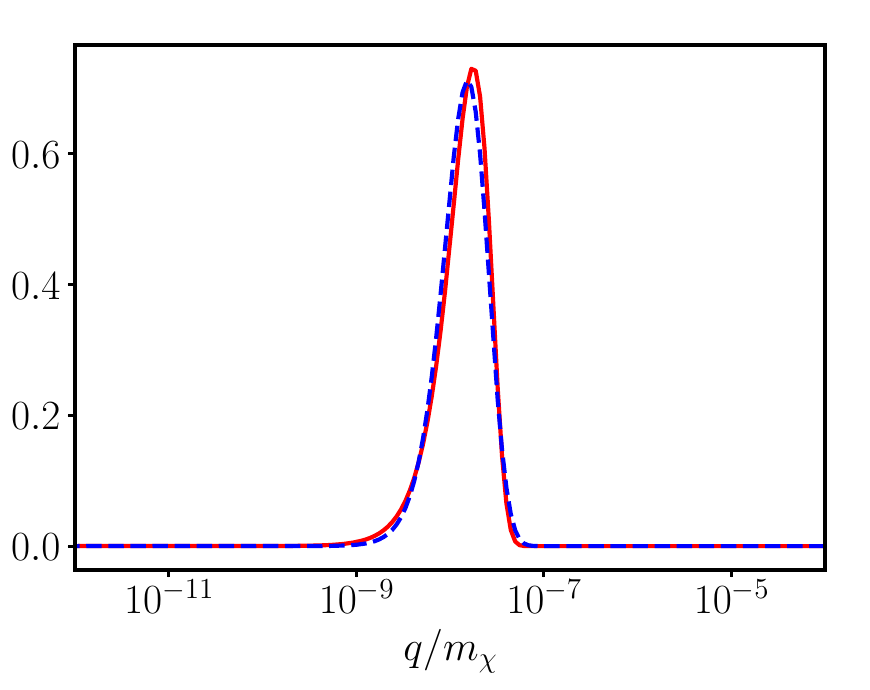} 
        \caption{$m_{\chi}=1\,\mathrm{GeV}$}
    \end{subfigure}
    \hfill 
    \begin{subfigure}[b]{0.32\textwidth}
        \centering
        \includegraphics[height=4cm]{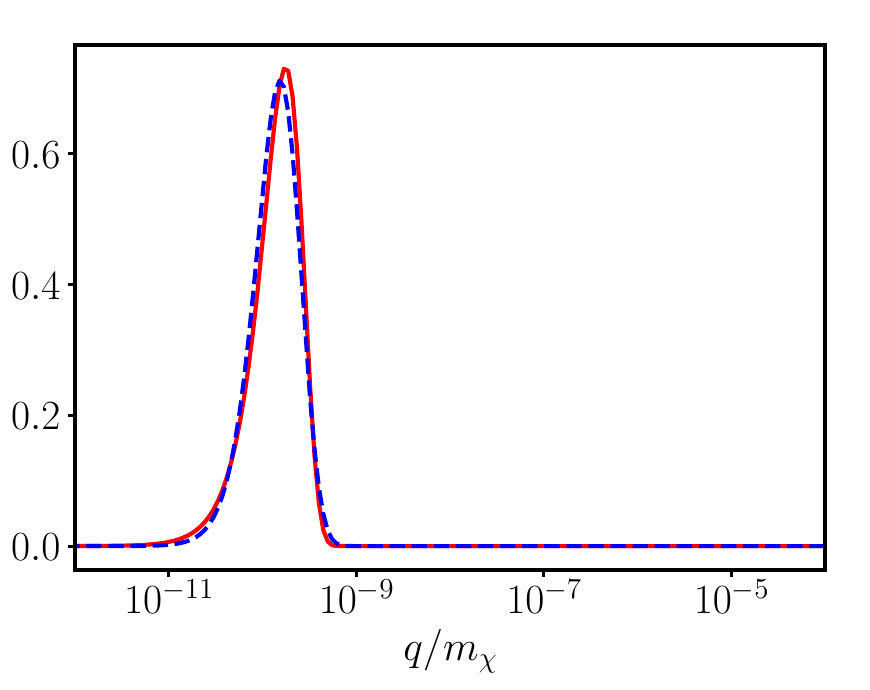} 
        \caption{$m_{\chi}=100\,\mathrm{GeV}$}
    \end{subfigure}
    \hfill 
    \begin{subfigure}[b]{0.32\textwidth}
        \centering
        \includegraphics[height=4cm]{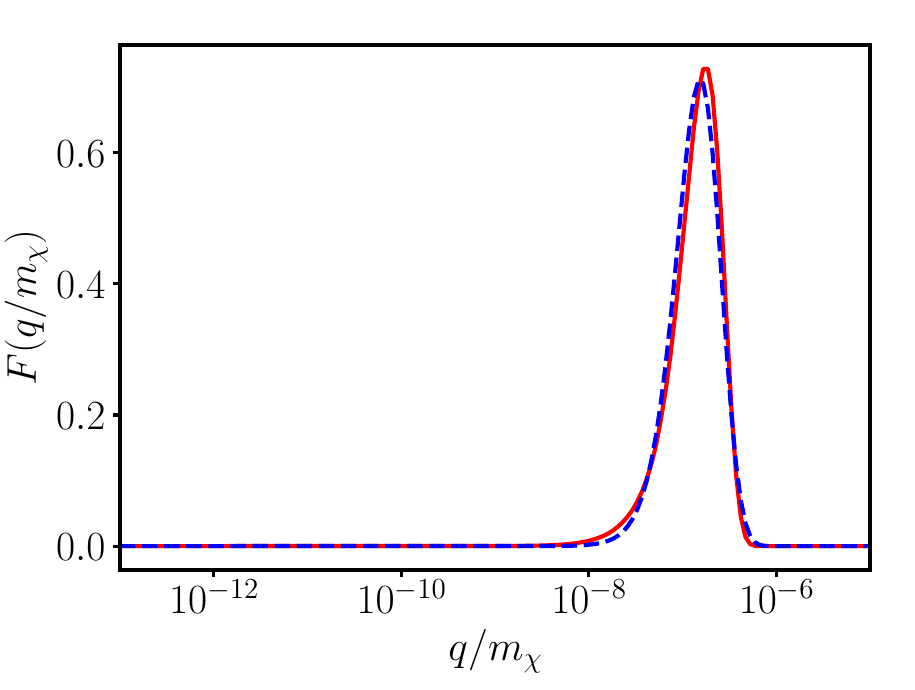} 
        \caption{$m_{\chi}=0.01\,\mathrm{GeV}$}
    \end{subfigure}
    \hfill 
    \begin{subfigure}[b]{0.30\textwidth}
        \centering
        \includegraphics[height=4cm]{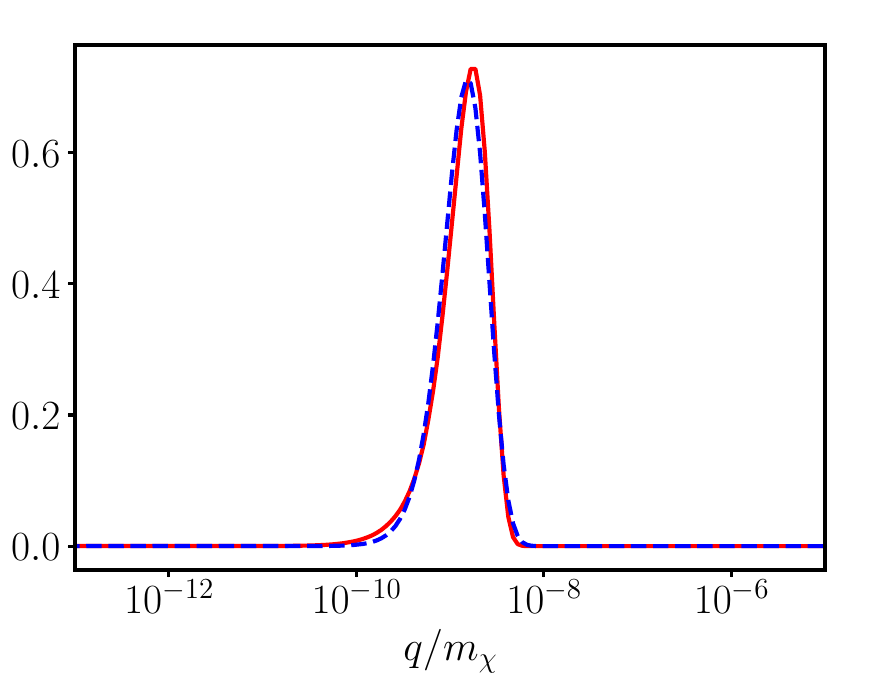} 
        \caption{$m_{\chi}=1\,\mathrm{GeV}$}
    \end{subfigure}
    \hfill 
    \begin{subfigure}[b]{0.32\textwidth}
        \centering
        \includegraphics[height=4cm]{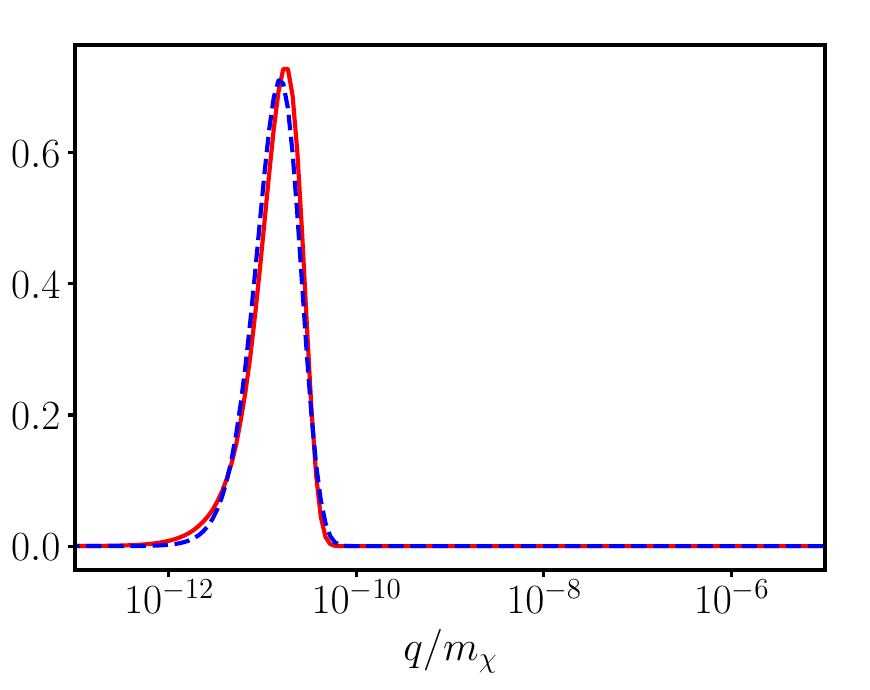} 
        \caption{$m_{\chi}=100\,\mathrm{GeV}$}
    \end{subfigure}

    \caption{Comparison of the normalized unperturbed background comoving phase-space distributions between this model (red solid lines) and the equivalent thermal WDM model (blue dashed lines). The horizontal axis represents the dimensionless comoving momentum $q/m_\chi$. The masses of the parent particle and the associated decay product are fixed at $m_\phi = 1000\,\mathrm{GeV}$ and $m_N = 1\,\mathrm{GeV}$ across all panels. The upper, middle, and lower rows correspond to coupling constants of $y_{\mathrm{DS}} = 10^{-13}$, $10^{-12}$, and $10^{-11}$, respectively. The left, center, and right columns represent DM masses of $m_\chi = 0.01\,\mathrm{GeV}$, $1\,\mathrm{GeV}$, and $100\,\mathrm{GeV}$, respectively.}
    \label{fig:wdm_approximation1}
\end{figure}

Fig.~\ref{fig:wdm_approximation1} compares the unperturbed background comoving phase-space distributions of the DM model with those of the equivalent thermal WDM model. 
In our analysis, we fix $m_\phi = 1000\,\mathrm{GeV}$ and $m_N = 1\,\mathrm{GeV}$. We then vary the DM mass $m_\chi$ for three benchmark couplings: $y_{\mathrm{DS}} = 10^{-13}$, $10^{-12}$, and $10^{-11}$. The largest coupling falls outside our considered parameter space of the decay-dominated 
DM production scenario, but we include it purely for illustration. Across all these cases, we observe remarkable consistency in the distribution shapes and peak positions. The models also yield identical $\langle v^2 \rangle$ values and consistent distribution widths. Furthermore, the precise impact of this equivalent mapping on the small-scale structure suppression must be quantified using the linear matter power spectrum, as detailed in section~\ref{subsec:transfer}.

\subsection{Linear Transfer Function}
\label{subsec:transfer}

The suppression of the small-scale matter power spectrum is directly quantified by the transfer function $T_X(k)$, defined as
\begin{equation}
    P_X(k) = P_{\mathrm{CDM}}(k) T_X^2(k) \,,
\end{equation}
where $P_X(k)$ and $P_{\mathrm{CDM}}(k)$ are the linear matter power spectra for a specific DM $X$ and the standard Cold Dark Matter (CDM), respectively, and $k$ is the comoving wavenumber. 

\begin{figure}[htbp]
    \centering
    \begin{subfigure}[b]{0.32\textwidth}
        \centering
        \includegraphics[height=4cm]{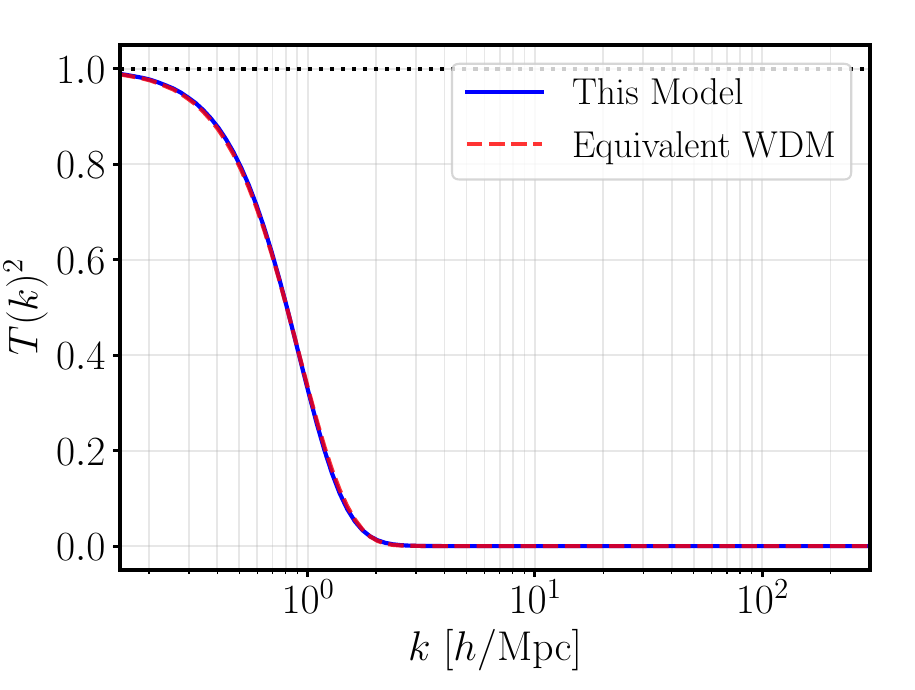} 
        \caption{$m_\chi=0.01\,\mathrm{GeV}$}
    \end{subfigure}
    \hfill 
    \begin{subfigure}[b]{0.30\textwidth}
        \centering
        \includegraphics[height=4cm]{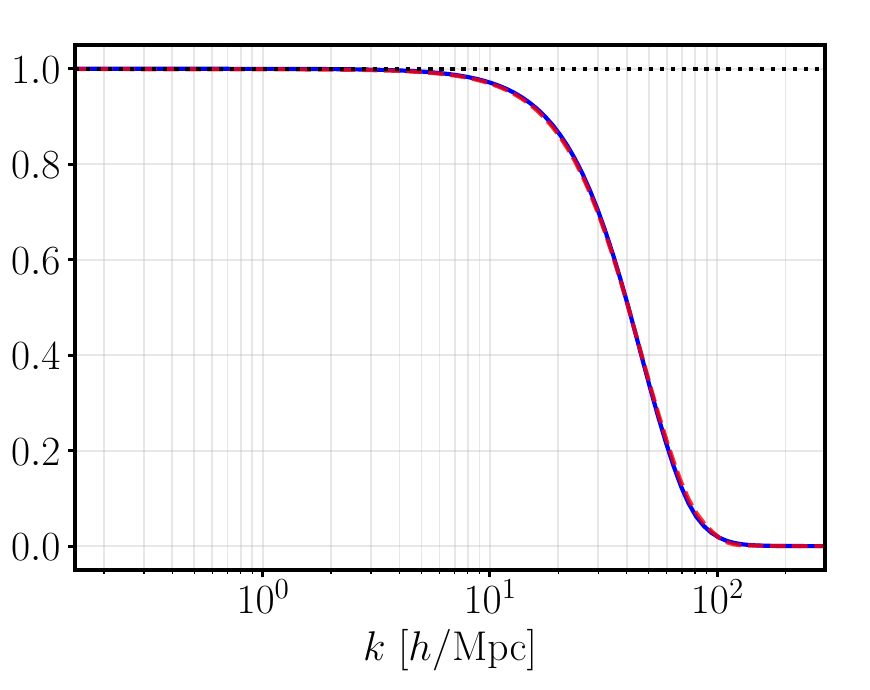} 
        \caption{$m_\chi=1\,\mathrm{GeV}$}
    \end{subfigure}
    \hfill    
     \begin{subfigure}[b]{0.32\textwidth}
        \centering
        \includegraphics[height=4cm]{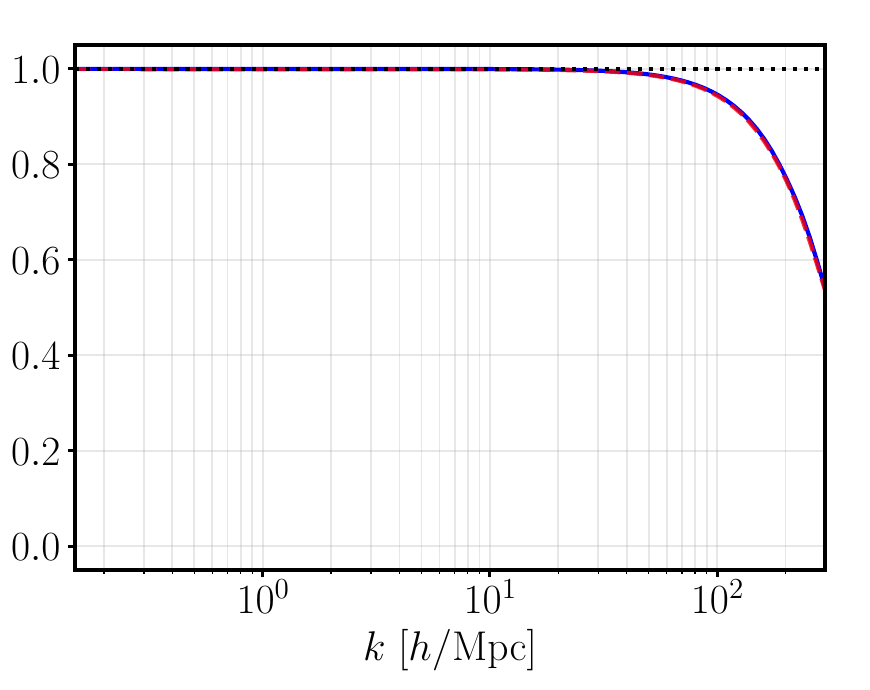} 
        \caption{$m_\chi=10\,\mathrm{GeV}$}
    \end{subfigure}

    \caption{Comparison of the numerical transfer functions between our model and its equivalent WDM counterpart, obtained from \texttt{CLASS} numerical calculations. The chosen DM masses $m_\chi \in \{0.01, 1, 10\}\,\mathrm{GeV}$ correspond to effective thermal masses of $m_{\mathrm{eff}} \simeq 0.1$, $5$, and $25\,\mathrm{keV}$, respectively. Other parameters are fixed at $m_\phi = 1000\,\mathrm{GeV}$, $y_{\mathrm{DS}} = 10^{-12}$, and $m_N = 1\,\mathrm{GeV}$.}
    \label{fig:wdm_approximation2}
\end{figure}

To validate the kinematic mapping established in the previous section, we first directly compare the numerical transfer functions. The linear matter power spectra for both the decay model and its equivalent thermal WDM counterpart are generated using the Boltzmann code \texttt{CLASS}~\cite{Blas:2011rf, Lesgourgues:2011re}, assuming the Planck 2018 best-fit cosmological parameters~\cite{Planck2018}.
As demonstrated in Fig.~\ref{fig:wdm_approximation2}, the numerical transfer functions for both models show a high degree of agreement. Here, we fix the parameters at $m_\phi = 1000\,\mathrm{GeV}$, $y_{\mathrm{DS}} = 10^{-12}$, and $m_N = 1\,\mathrm{GeV}$, while the chosen DM masses $m_\chi \in \{0.01, 1, 10\}\,\mathrm{GeV}$ correspond to effective thermal masses of $m_{\mathrm{eff}} \simeq 0.1$, $5$, and $25\,\mathrm{keV}$, respectively.

To strictly quantify this agreement,  we notice that the transfer function can be parameterized by a dimensionless exponent 
$\mu = 1.12$ and a breaking scale $\alpha_X$~\cite{FISW:2021, Murgia:2017}:
\begin{equation}
    T_X(k) = \left[ 1 + (\alpha_X k)^{2\mu} \right]^{-5/\mu} \,.
    \label{eq:transfer_func}
\end{equation}
We extract the breaking scale $\alpha_X$ for both our $\chi$ DM and the equivalent WDM approximation by performing a single-parameter fit to the numerical transfer functions using Eq.~(\ref{eq:transfer_func}), keeping the exponent fixed at $\mu = 1.12$. 
As summarized in Table~\ref{tab:alpha_comparison1}, the extracted $\alpha_\chi$ values for the decay model  and its equivalent thermal WDM counterpart $\alpha_{\mathrm{WDM}}$ maintain a relative error within $1\%$. Furthermore, we have performed analogous calculations across a broader parameter space, and the results remain robust, with relative errors consistently bounded within $2\%$.

\begin{table}[htbp]
    \centering
    \renewcommand{\arraystretch}{1.3} 
    \begin{tabular}{@{} c c c c c @{}}
        \hline\hline
        $m_\chi$ [GeV] & $m_{\mathrm{eff}}$ [keV] & $\alpha_{\chi}$ [$h^{-1}\,\mathrm{Mpc}$] & $\alpha_{\mathrm{WDM}}$ [$h^{-1}\,\mathrm{Mpc}$] & Relative Error \\
        \hline
        $0.01$ & $0.1$ & $0.404044$ & $0.402175$ & $0.5\%$ \\
        $1$    & $5$   & $0.007884$ & $0.007819$ & $0.8\%$ \\
        $10$   & $25$  & $0.001031$ & $0.001039$ & $0.7\%$ \\
        \hline\hline
    \end{tabular}
    \caption{Comparison of the extracted breaking scales $\alpha_X$ for the decay model ($\alpha_{\chi}$) and its equivalent thermal WDM counterpart ($\alpha_{\mathrm{WDM}}$). Other parameters are identical to those in Fig.~\ref{fig:wdm_approximation2}.}
    \label{tab:alpha_comparison1}
\end{table}

Despite the high precision of the full numerical calculation, evaluating the transfer functions across the entire parameter space is computationally expensive. Therefore, to improve computational efficiency, we adopt an analytical approximation.
For thermal WDM, the breaking scale can be analytically expressed as a function of the WDM mass~\cite{FISW:2021}, which is given by:
\begin{equation}
    \alpha_{\mathrm{WDM}} = 0.045 \left(\frac{m_{\mathrm{WDM}}}{1\,\mathrm{keV}}\right)^{-1.11} \left(\frac{\Omega_{\mathrm{WDM}}}{0.25}\right)^{0.11} \left(\frac{h}{0.7}\right)^{1.22} h^{-1}\,\mathrm{Mpc} \,.
    \label{eq:alpha_WDM}
\end{equation}
Within the mass range of $3$\,--\,$5\,\mathrm{keV}$, the relative error of this analytical prefactor is known to be less than $1.5\%$~\cite{FISW:2021}.

\begin{figure}[htbp]
    \centering
    \begin{subfigure}[b]{0.35\textwidth}
        \centering
        \includegraphics[height=4cm]{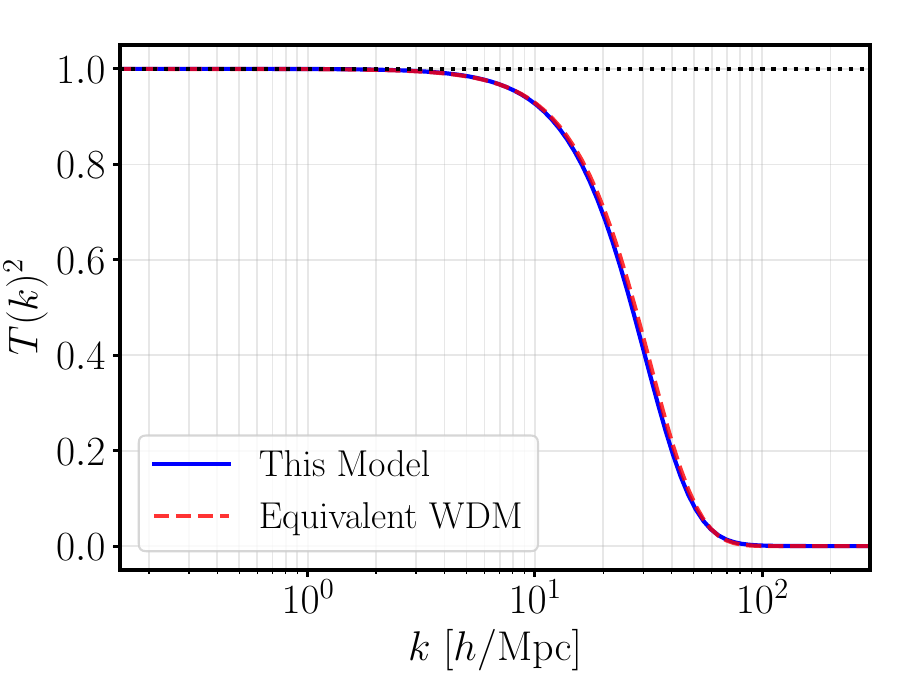} 
        \caption{$m_\chi=0.62\,\mathrm{GeV}$}
    \end{subfigure}
    \begin{subfigure}[b]{0.35\textwidth}
        \centering
        \includegraphics[height=4cm]{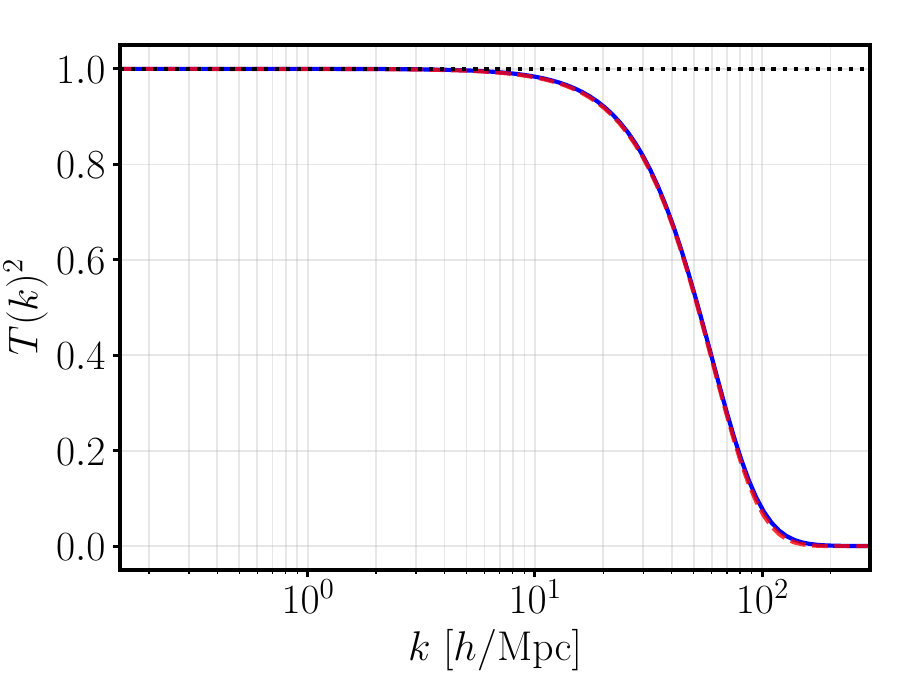} 
        \caption{$m_\chi=1.33\,\mathrm{GeV}$}
    \end{subfigure}

    \caption{Comparison of the transfer function of our model against the equivalent WDM predictions derived from Eq.~(\ref{eq:alpha_WDM}). 
Other parameters are fixed at $m_\phi = 1000\,\mathrm{GeV}$, $y_{\mathrm{DS}} = 10^{-12}$, and $m_N = 1\,\mathrm{GeV}$. The left and right panels correspond to $m_\chi = 0.62$ and $1.33\,\mathrm{GeV}$, which yield effective WDM masses of $m_{\mathrm{eff}} \simeq 3.2$ and $5.7\,\mathrm{keV}$, respectively.}
    \label{fig:transfer_fit}
\end{figure}

To validate this analytical formula for our model, Fig.~\ref{fig:transfer_fit} compares the transfer function of $\chi$ DM obtained in our model against the predictions of the equivalent WDM with $\alpha_{\mathrm{WDM}}$ calculated using Eq.~(\ref{eq:alpha_WDM}). We specifically examine the parameters yielding $m_{\mathrm{eff}} \simeq 3.2\,\mathrm{keV}$ and $5.7\,\mathrm{keV}$, as these correspond to the observational boundaries adopted in our subsequent analysis in section~\ref{sec:results}. For these benchmark values, the relative error in $\alpha_X$ remains within $2\%$, as detailed in Table~\ref{tab:alpha_comparison2}. This conclusion remains robust across different parameter combinations that yield the same $m_{\mathrm{eff}}$.

\begin{table}[htbp]
    \centering
    \renewcommand{\arraystretch}{1.3} 
    \begin{tabular}{@{} c c c c c @{}}
        \hline\hline
        $m_\chi$ [GeV] & $m_{\mathrm{eff}}$ [keV] & $\alpha_{\chi}$ [$h^{-1}\,\mathrm{Mpc}$] & $\alpha_{\mathrm{WDM}}$ [$h^{-1}\,\mathrm{Mpc}$] & Relative Error \\
        \hline
        $0.62$ & $3.2$ & $0.012003$ & $0.011770$ & $2\%$ \\
        $1.33$    & $5.7$   & $0.006116$ & $0.006235$ & $2\%$ \\
        \hline\hline
    \end{tabular}
    \caption{Comparison of the extracted breaking scales $\alpha_X$ for the decay model ($\alpha_{\chi}$) and its equivalent thermal WDM counterpart ($\alpha_{\mathrm{WDM}}$). Other parameters are identical to those in Fig.~\ref{fig:transfer_fit}.}
    \label{tab:alpha_comparison2}
\end{table}

The agreement between the transfer function of $\chi$ DM and the prediction of the equivalent WDM with $\alpha_{\mathrm{WDM}}$ calculated  
using Eq.~(\ref{eq:alpha_WDM}) 
suggests that we can obtain an equivalent WDM approximation in another way, independent of the matching method using velocity dispersion presented
in (\ref{eq:momentum}). Namely, we can fit the transfer function of $\chi$ DM in our model yielding a breaking scale $\alpha_\chi$.
Taking this $\alpha_\chi$ as $\alpha_{\mathrm{WDM}}$ and inserting this value into Eq.~(\ref{eq:alpha_WDM}), we can 
invert this expression and obtain an effective mass $m_{\mathrm{eff}}$ of the equivalent WDM approximation. This provides another way 
to project Lyman-$\alpha$ constraint onto our model without $N$-body simulations, as will be detailed in the next section.

\section{Lyman-$\alpha$ forest constraint on the model}
\label{sec:results}

In this section, we study the constraint on the DM model from the Lyman-$\alpha$ forest observations. 
We then combine limits from the Lyman-$\alpha$ forest observations and the BBN constraint.
Requiring the $\chi$ particles to fully account for the observed DM relic abundance uniquely determines the initial coupling $\lambda_{H\phi}$. Consequently, our phenomenological study focuses on four primary free parameters: the parent particle mass $m_\phi$, the DM mass $m_\chi$, the decay coupling $y_{\mathrm{DS}}$, and the neutrino mass $m_N$. 

Recent lower bounds on the WDM mass derived from various observational probes are summarized in Table~\ref{tab:wdm_constraints}. For the subsequent analysis, we adopt the Lyman-$\alpha$ forest bound ($m_{\mathrm{WDM}} \geq 5.7\,\mathrm{keV}$) as our fiducial constraint, while treating the limit derived from UV Luminosity Functions ($m_{\mathrm{WDM}} \geq 3.2\,\mathrm{keV}$) as a conservative alternative. We emphasize that this constraint method applies to all the listed lower bounds.

\begin{table}[htbp]
    \centering
    \renewcommand{\arraystretch}{1.3} 
    \begin{tabular}{@{} l l p{5.2cm} @{}}
        \hline\hline
        \textbf{Observational Probe} & \textbf{Lower Bound} ($m_{\mathrm{WDM}}$) & \textbf{Data Source} \\
        \hline
        Lyman-$\alpha$ forest & $\geq 5.7\,\mathrm{keV}$ & High-resolution quasar spectra (HIRES and UVES)~\cite{Irsic2024} \\
        Milky Way Satellite Counts & $\geq 5.9$\,--\,$6.2\,\mathrm{keV}$ & Milky Way and M31 satellite galaxy census (e.g., DES, PS1 and PAndAS)~\cite{Nadler:2024_COZMIC, Liu:2025_MW} \\
        Stellar Streams & $\geq 3.6\,\mathrm{keV}$ & Gaia observations of GD-1 and Palomar 5 streams~\cite{Banik:2021} \\
        UV Luminosity Functions & $\geq 3.2\,\mathrm{keV}$ & High-redshift galaxy surveys from JWST~\cite{Liu:2024_JWST} \\
        \hline\hline
    \end{tabular}
    \caption{Summary of recent observational lower bounds on the mass of thermal relic WDM.}
    \label{tab:wdm_constraints}
\end{table}

As discussed in ref.~\cite{ChengYu:2021}, the late-time decay of the dark sector WIMP $\phi$ can inject substantial energy into the thermal bath. To avoid disrupting the primordial light element abundances, the lifetime of $\phi$ is strictly constrained by BBN. Specifically, we adopt two BBN limits on the decay lifetime: a conservative bound of $\tau \leq 0.7\,\mathrm{s}$ (corresponding to a thermal bath temperature $T \gtrsim 1\,\mathrm{MeV}$) and a stringent bound of $\tau \leq 0.07\,\mathrm{s}$ (corresponding to $T \gtrsim 3\,\mathrm{MeV}$). 

We first explore the parameter space using the highly efficient velocity dispersion method, followed by an explicit verification through transfer function fitting. To demonstrate the constraining power of the Lyman-$\alpha$ forest, we adopt the stringent Lyman-$\alpha$ limit alongside the conservative BBN bound as our comparative benchmarks.

\subsection{Velocity Dispersion Constraints}
\label{subsec:velocity_constraints}

To systematically extract constraints using the velocity dispersion approach, we implement the following step-by-step procedure for a given set of model parameters ($m_\phi, m_\chi, m_N, y_{\mathrm{DS}}$):
\begin{enumerate}
    \item Evaluate the normalized unperturbed background comoving phase-space distribution $F(q)$ of the $\chi$ DM particles by numerically solving the Boltzmann eq.~(\ref{eq:Cchi_F}). Using this non-thermal distribution, we compute the number density $n_{\mathrm{\chi}}$ and the mean squared comoving momentum $\langle q^2 \rangle_{\mathrm{\chi}}$.
    \item Using the matching conditions established in eqs.~(\ref{eq:abundance}) and (\ref{eq:momentum}), we map these derived quantities to an equivalent thermal WDM model. By solving the two equations, we obtain an effective temperature $T_{\mathrm{eff}}$ and an effective mass $m_{\mathrm{eff}}$.
    \item This $m_{\mathrm{eff}}$ is directly compared against the established observational limits. If $m_{\mathrm{eff}} < m_{\mathrm{WDM}}^{\mathrm{limit}}$ (e.g., $5.7\,\mathrm{keV}$ for the stringent Lyman-$\alpha$ bound), the DM velocity dispersion is overly ``hot''. Such parameter points excessively suppress small-scale structures and are therefore excluded.
\end{enumerate}

\begin{figure}[tbp]
    \centering
    \begin{subfigure}[b]{0.32\textwidth}
        \centering
        \includegraphics[height=3.8cm]{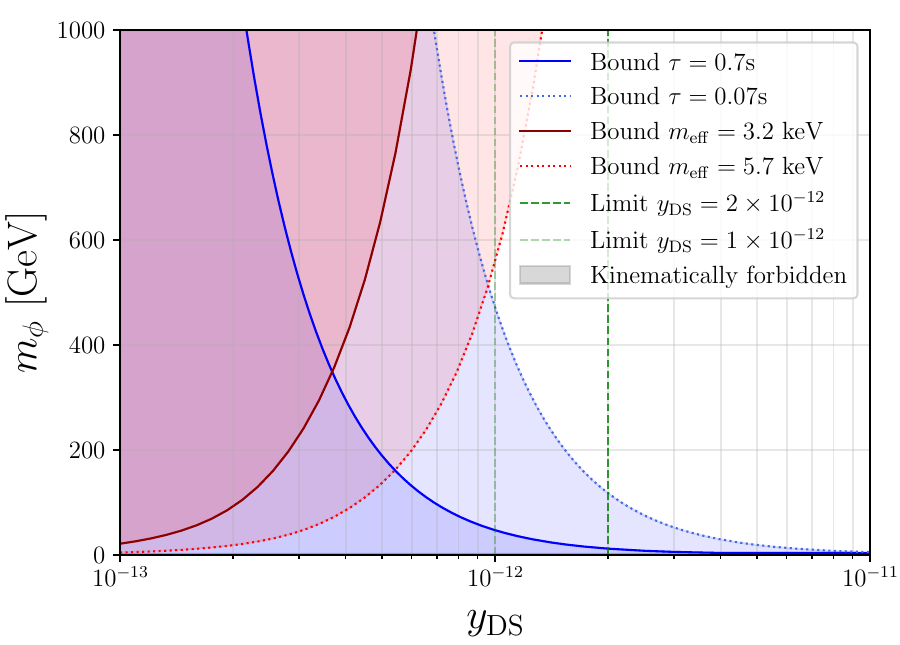} 
        \caption{$m_{\chi}=1\,\mathrm{GeV}$}
    \end{subfigure}
    \hfill 
    \begin{subfigure}[b]{0.30\textwidth}
        \centering
        \includegraphics[height=3.8cm]{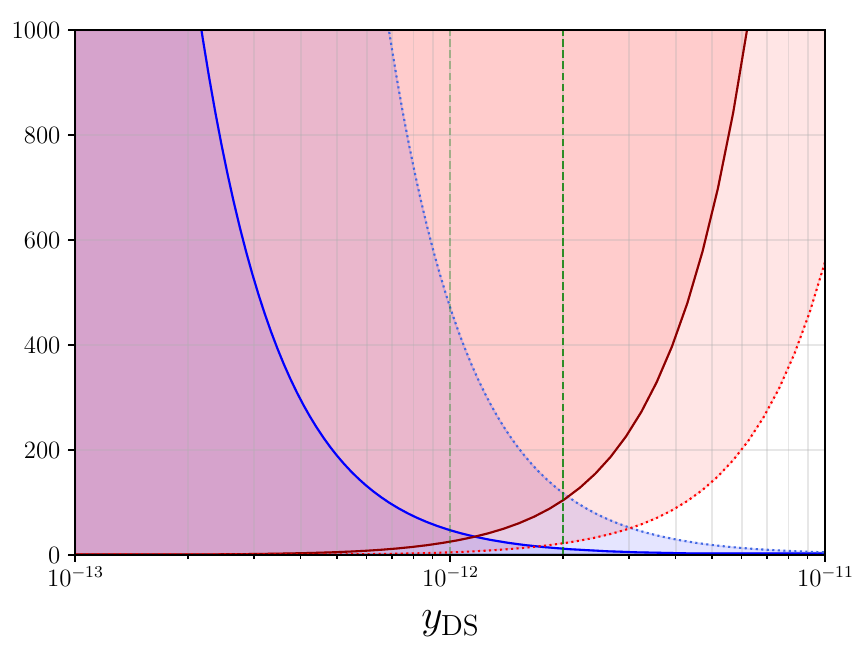} 
        \caption{$m_{\chi}=0.1\,\mathrm{GeV}$}
    \end{subfigure}
    \hfill    
     \begin{subfigure}[b]{0.32\textwidth}
        \centering
        \includegraphics[height=3.8cm]{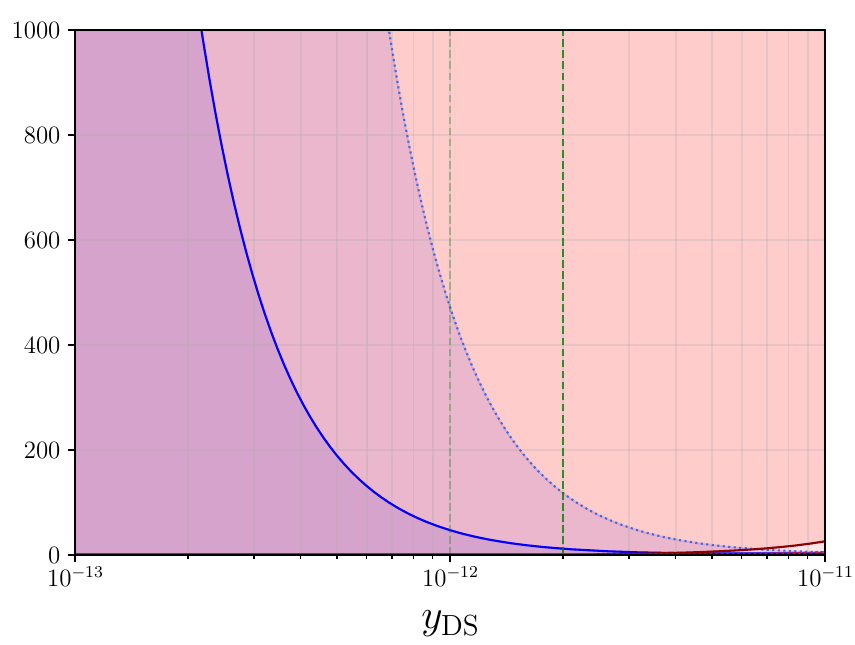} 
        \caption{$m_{\chi}=0.01\,\mathrm{GeV}$}
    \end{subfigure}

    \caption{Constraints on the $(y_{\mathrm{DS}}, m_\phi)$ parameter space for varying DM masses $m_\chi \in \{1, 0.1, 0.01\}\,\mathrm{GeV}$ with a fixed neutrino mass $m_N = 1\,\mathrm{GeV}$. The red solid and dotted curves correspond to the boundaries given by the equivalent WDM masses $m_{\mathrm{eff}} = 3.2\,\mathrm{keV}$ and $5.7\,\mathrm{keV}$, respectively. The blue solid and dotted lines denote the conservative ($\tau = 0.7\,\mathrm{s}$) and stringent ($\tau = 0.07\,\mathrm{s}$) BBN limits, respectively. 
The shaded regions to the left of the corresponding lines are excluded by the corresponding constraints.
The green dashed and light green dashed lines indicate $y_{\mathrm{DS}} = 2 \times 10^{-12}$ ($90\%$ decay production) and $y_{\mathrm{DS}} = 1 \times 10^{-12}$ ($95\%$ decay production), respectively. 
}
    \label{fig:results1}
\end{figure}

Applying this numerical workflow to mediator masses within the typical WIMP scale ($1\,\mathrm{GeV} < m_\phi < 1\,\mathrm{TeV}$), we derive the parameter space constraints illustrated in Fig.~\ref{fig:results1}. Specifically,
we present the allowed regions in the $(y_{\mathrm{DS}}, m_\phi)$ plane for fixed values of $m_\chi$ or $m_N$. 
In these panels, the red lines represent the boundaries determined by the equivalent WDM velocity dispersion method, corresponding to the Lyman-$\alpha$ limits of $m_{\mathrm{eff}} = 5.7\,\mathrm{keV}$ (stringent) and $3.2\,\mathrm{keV}$ (conservative). Meanwhile, the blue lines denote the BBN constraints on the mediator lifetime ($\tau \leq 0.07\,\mathrm{s}$ and $0.7\,\mathrm{s}$). 

These two bounds provide highly complementary constraints because they exhibit opposite dependencies on the underlying model parameters.
For the velocity dispersion constraints (red lines), the underlying physics is straightforward. A larger parent mass $m_\phi$ imparts a greater initial momentum to the $\chi$ particles during decay. To suppress the late-time velocity dispersion of the DM distribution, these particles must undergo more cosmic redshifting. This requires the $\phi$ particles to decay earlier. An earlier decay corresponds to a shorter lifetime $\tau$. Consequently, a larger decay coupling $y_{\mathrm{DS}}$ is required. Therefore, a larger $m_\phi$ must be compensated by a larger $y_{\mathrm{DS}}$ for this red line of constraint.
Conversely, the BBN limits (blue lines) are dictated solely by the decay rate $\Gamma_\phi$.
A larger $m_\phi$ can give rise to a larger phase space for $\phi$ decay and a larger $\Gamma_\phi$ which can results in a $\phi$ decay at earlier times.
To saturate the BBN bound, smaller $y_{\mathrm{DS}}$ is required. So we have the inverse correlation of $m_\phi$ and $y_{\mathrm{DS}}$ of the BBN line in the figure.

Fig.~\ref{fig:results1} also illustrates the effect of decreasing $m_\chi$ with a fixed $m_N = 1\,\mathrm{GeV}$. 
Kinematically, a smaller $m_\chi$ yields a larger decay momentum, a larger Lorentz factor $\gamma$ and a larger late-time velocity dispersion. 
This enhanced free-streaming systematically tightens the Lyman-$\alpha$ bounds, pushing the red curves toward larger $y_{\mathrm{DS}}$. For DM masses in the GeV regime (panel a, $m_\chi = 1\,\mathrm{GeV}$), the velocity dispersion limits only marginally extend beyond the BBN constraints, excluding $y_{\mathrm{DS}}$ up to $\sim 1 \times 10^{-12}$. In this regime, the allowed mediator mass $m_\phi$ can still comfortably reach the upper scale of the typical WIMP window at $1000\,\mathrm{GeV}$.
Furthermore, unlike BBN, the Lyman-$\alpha$ bounds impose a strict upper ceiling on $m_\phi$ to prevent excessive initial decay momentum. This kinematic ceiling drops drastically as $m_\chi$ decreases. As shown in panel (b) for $m_\chi = 0.1\,\mathrm{GeV}$, the maximum allowed $m_\phi$ is capped at $\sim 500\,\mathrm{GeV}$. Ultimately, for $m_\chi = 0.01\,\mathrm{GeV}$ (panel c), this parameter space ceiling collapses entirely; even the most conservative Lyman-$\alpha$ and BBN limits merge to completely exclude the decay production scenario for this mass of $\chi$ DM.

\begin{figure}[tbp]
    \centering
    \begin{subfigure}[b]{0.32\textwidth}
        \centering
        \includegraphics[height=3.8cm]{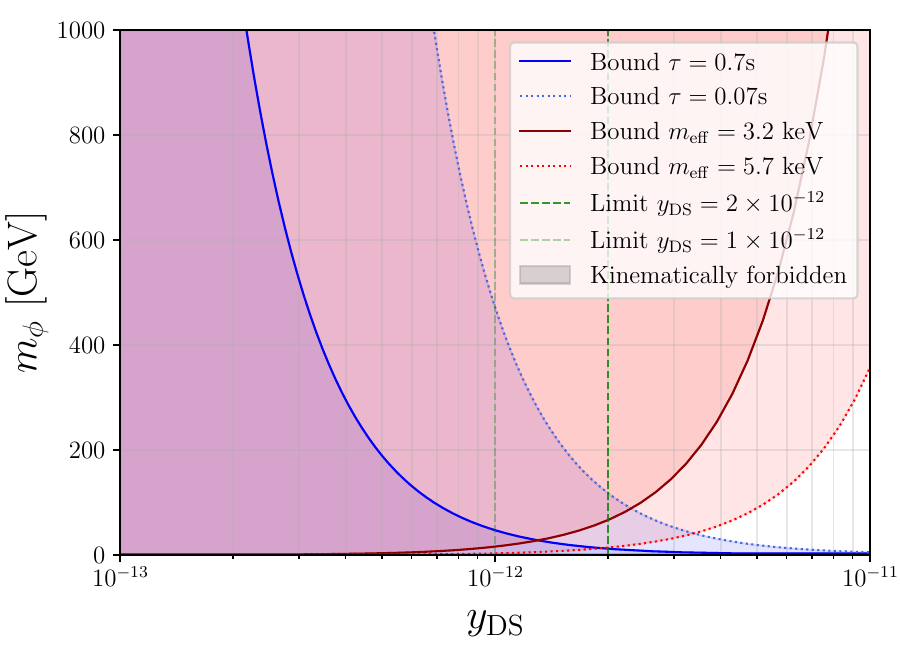} 
        \caption{$m_{N}=1\,\mathrm{GeV}$}
    \end{subfigure}
    \hfill     
    \begin{subfigure}[b]{0.30\textwidth}
        \centering
        \includegraphics[height=3.8cm]{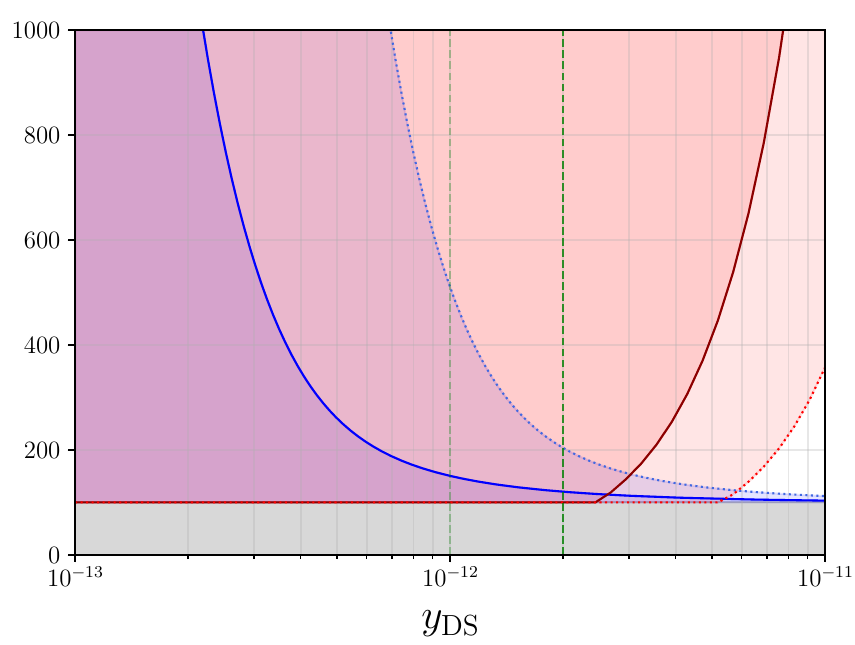} 
        \caption{$m_{N}=100\,\mathrm{GeV}$}
    \end{subfigure}
    \hfill     
    \begin{subfigure}[b]{0.32\textwidth}
        \centering
        \includegraphics[height=3.8cm]{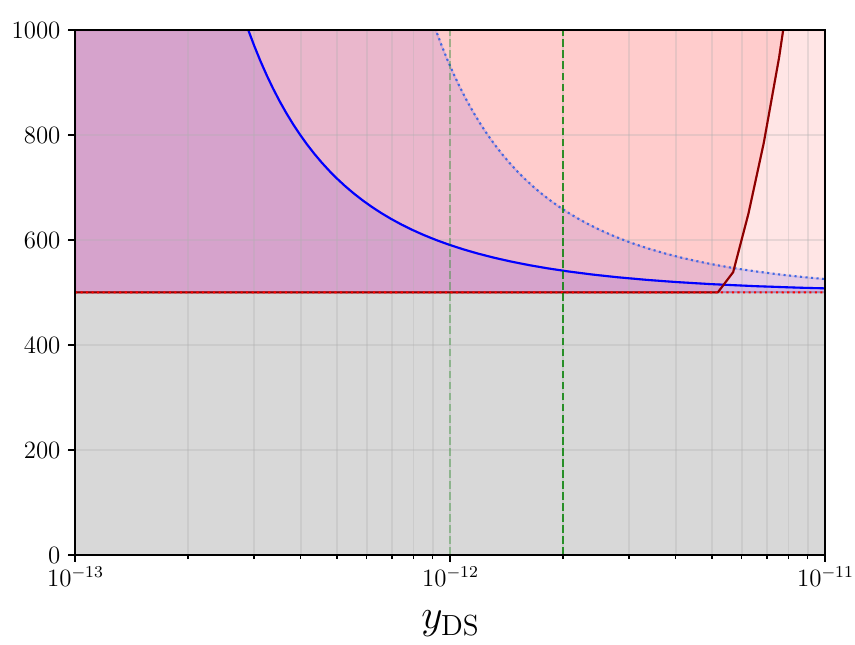} 
        \caption{$m_{N}=500\,\mathrm{GeV}$}
    \end{subfigure}
    \hfill
    \begin{subfigure}[b]{0.32\textwidth}
        \centering
        \includegraphics[height=3.8cm]{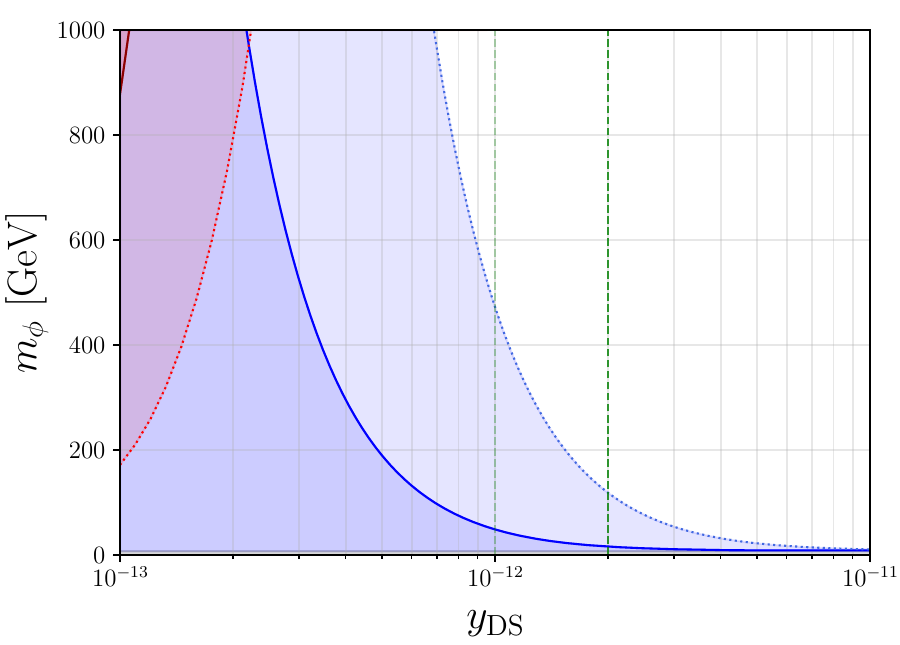} 
        \caption{$m_{N}=1\,\mathrm{GeV}$}
    \end{subfigure}
    \hfill 
    \begin{subfigure}[b]{0.30\textwidth}
        \centering
        \includegraphics[height=3.8cm]{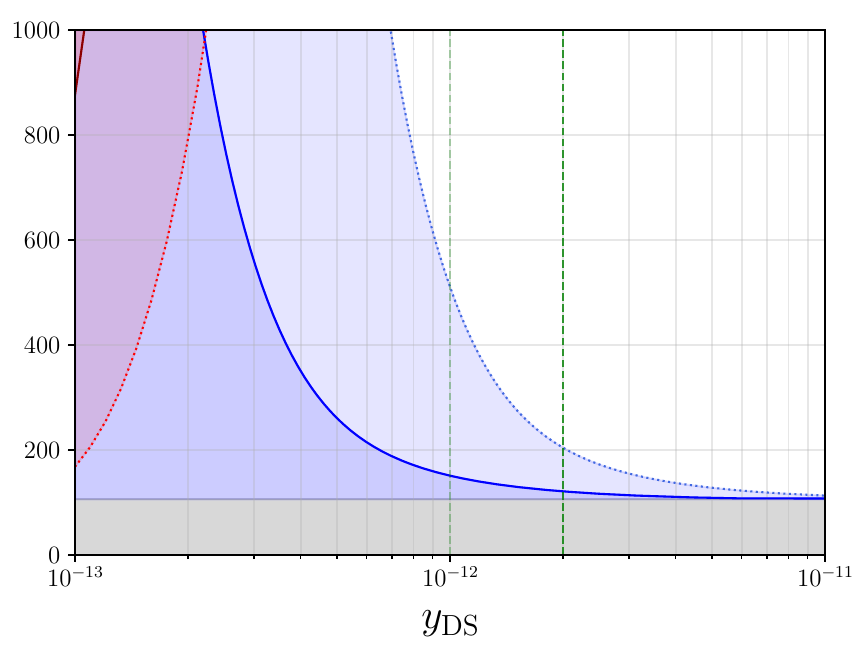} 
        \caption{$m_{N}=100\,\mathrm{GeV}$}
    \end{subfigure}
    \hfill    
     \begin{subfigure}[b]{0.32\textwidth}
        \centering
        \includegraphics[height=3.8cm]{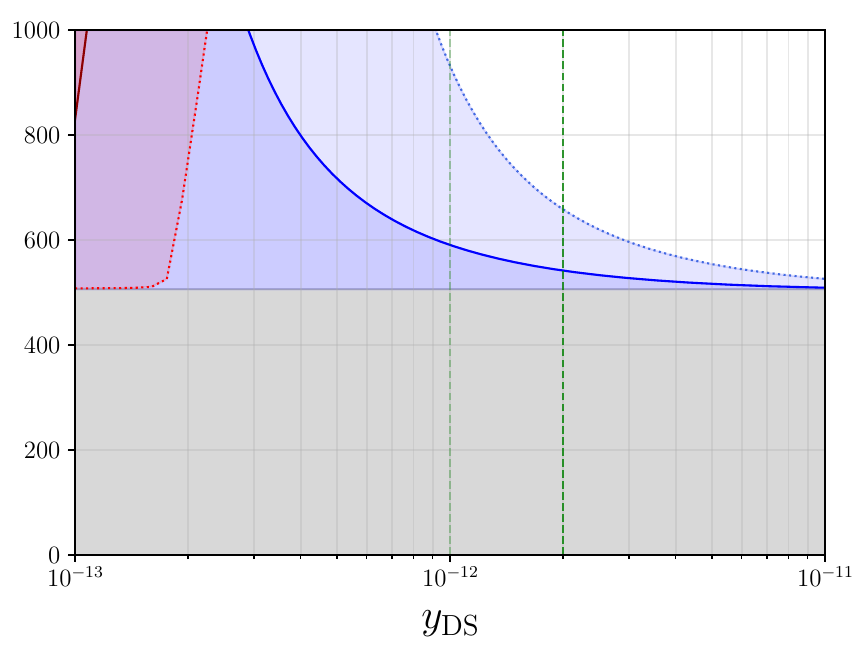} 
        \caption{$m_{N}=500\,\mathrm{GeV}$}
    \end{subfigure}

    \caption{Constraints on the $(y_{\mathrm{DS}}, m_\phi)$ parameter space for varying neutrino masses $m_N \in \{1, 100, 500\}\,\mathrm{GeV}$. The upper panels correspond to a fixed DM mass of $m_\chi = 0.08\,\mathrm{GeV}$, while the lower panels correspond to $m_\chi = 6\,\mathrm{GeV}$. The line styles and shaded regions have the same meanings as in Fig.~\ref{fig:results1}. The grey shaded regions at the bottom represent the kinematically forbidden parameter space ($m_\phi \le m_\chi + m_N$).
}
    \label{fig:results2}
\end{figure}

As shown in Fig.~\ref{fig:results2}, we vary the neutrino mass in the range $m_N \in \{1, 100, 500\}\,\mathrm{GeV}$, the basic kinematic requirement $m_\phi > m_N + m_\chi$ imposes a strict baseline constraint across all panels.
In the upper panels ($m_\chi = 0.08\,\mathrm{GeV}$), the joint Lyman-$\alpha$ and BBN constraints in Fig.~\ref{fig:results2}(a) push the allowed region to $y_{\mathrm{DS}} \gtrsim 2 \times 10^{-12}$, entirely ruling out the decay production scenario. Increasing $m_N$ only further tightens these limits; as shown in Figs.~\ref{fig:results2}(b) and (c), the combined bounds robustly exclude DM masses $m_\chi \lesssim 0.08\,\mathrm{GeV}$.

In contrast, the lower panels of Fig.~\ref{fig:results2} present the parameter space constraints for a heavier DM scenario ($m_\chi = 6\,\mathrm{GeV}$) with varying masses $m_N \in \{1, 100, 500\}\,\mathrm{GeV}$. In contrast to the sub-GeV regime, the constraints here are overwhelmingly dominated by the BBN limits (blue curves), because a heavier DM particle possesses a much smaller velocity dispersion at late times. Consequently, the Lyman-$\alpha$ bounds (red curves) are weakened. Furthermore, increasing $m_N$ expands the kinematically forbidden region. 
We can see that GeV scale $\chi$ DM is allowed by all these constraints with $y_{\mathrm{DS}}\sim 10^{-12}$.

\begin{figure}[tbp]
    \centering
    \begin{subfigure}[b]{0.32\textwidth}
        \centering
        \includegraphics[height=3.8cm]{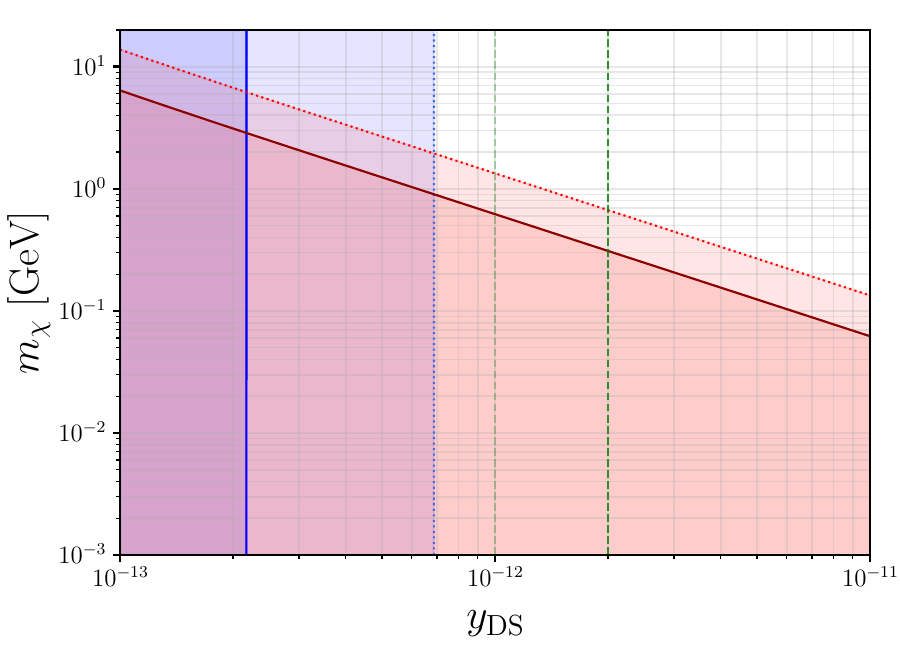} 
        \caption{$m_{N}=1\,\mathrm{GeV}$}
    \end{subfigure}
    \hfill 
    \begin{subfigure}[b]{0.30\textwidth}
        \centering
        \includegraphics[height=3.8cm]{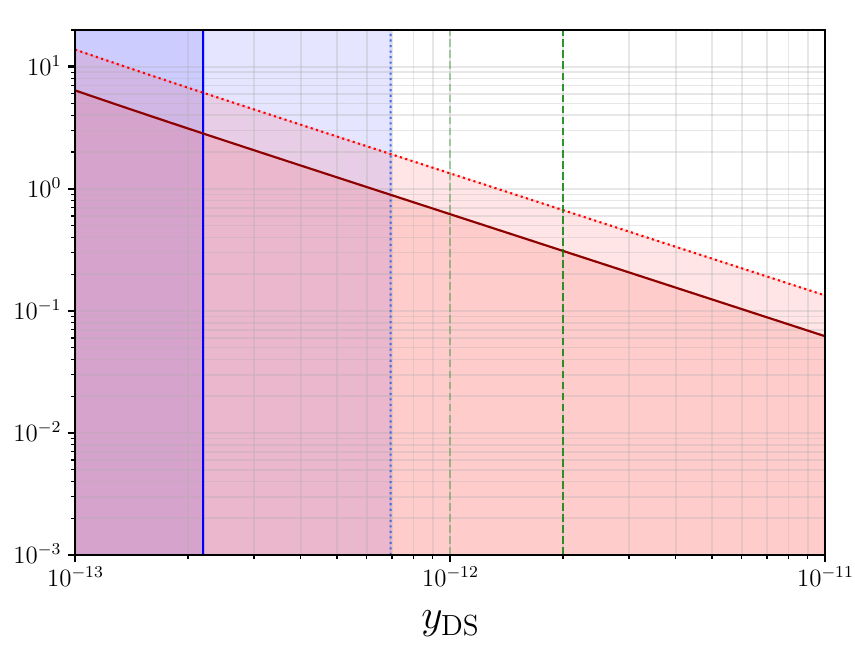} 
        \caption{$m_{N}=100\,\mathrm{GeV}$}
    \end{subfigure}
    \hfill    
     \begin{subfigure}[b]{0.32\textwidth}
        \centering
        \includegraphics[height=3.8cm]{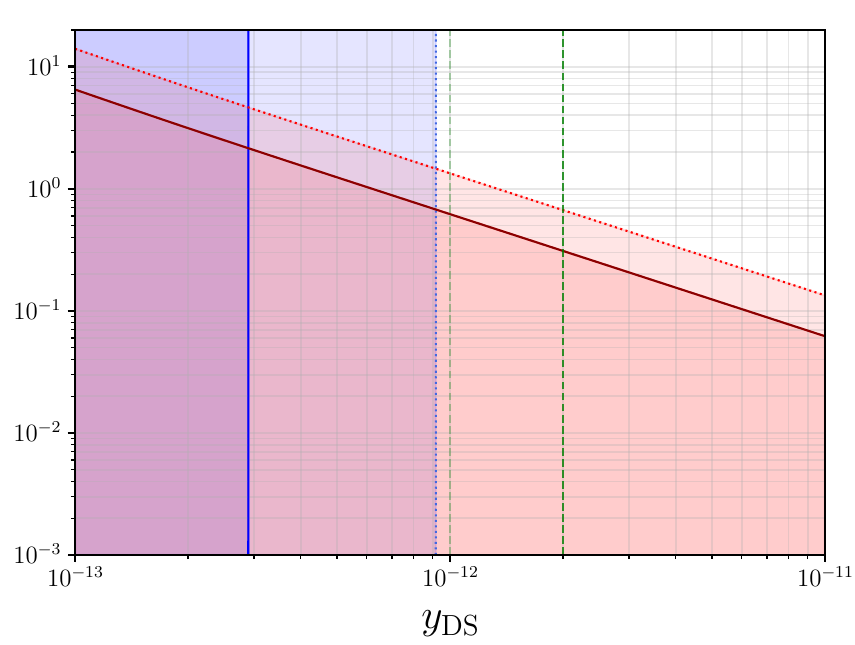} 
        \caption{$m_{N}=500\,\mathrm{GeV}$}
    \end{subfigure}
    \hfill     
    \begin{subfigure}[b]{0.32\textwidth}
        \centering
        \includegraphics[height=3.8cm]{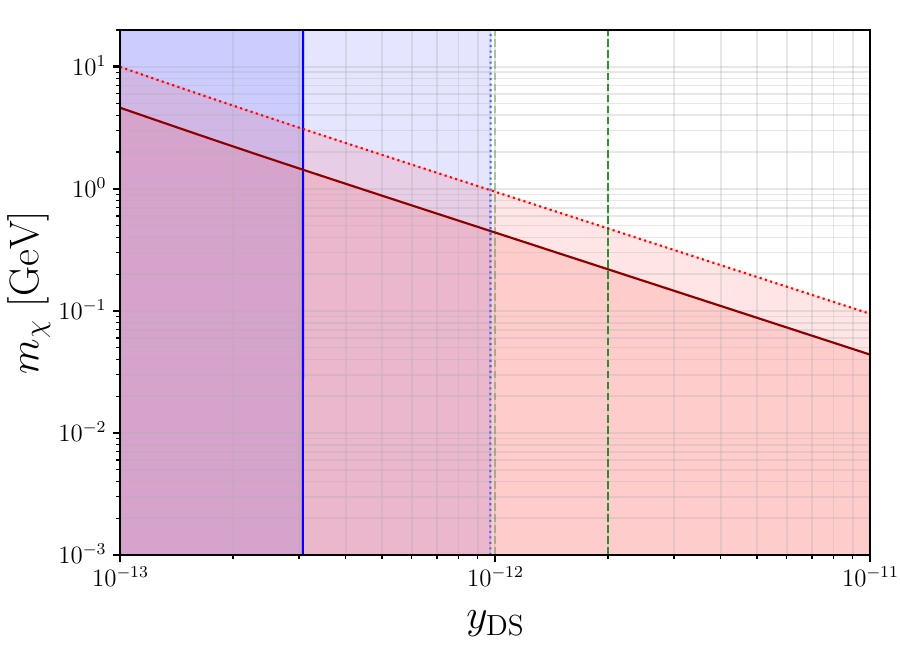} 
    \caption{$m_\phi=500\,\mathrm{GeV}$}
    \end{subfigure}
    \hfill     
    \begin{subfigure}[b]{0.30\textwidth}
        \centering
        \includegraphics[height=3.8cm]{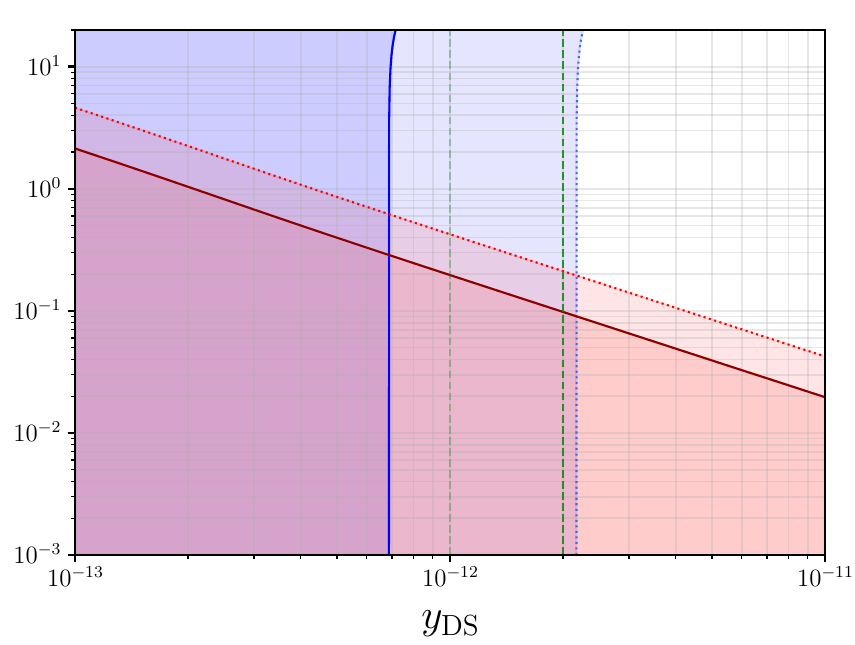} 
        \caption{$m_\phi=100\,\mathrm{GeV}$}
    \end{subfigure}
    \hfill     
    \begin{subfigure}[b]{0.32\textwidth}
        \centering
        \includegraphics[height=3.8cm]{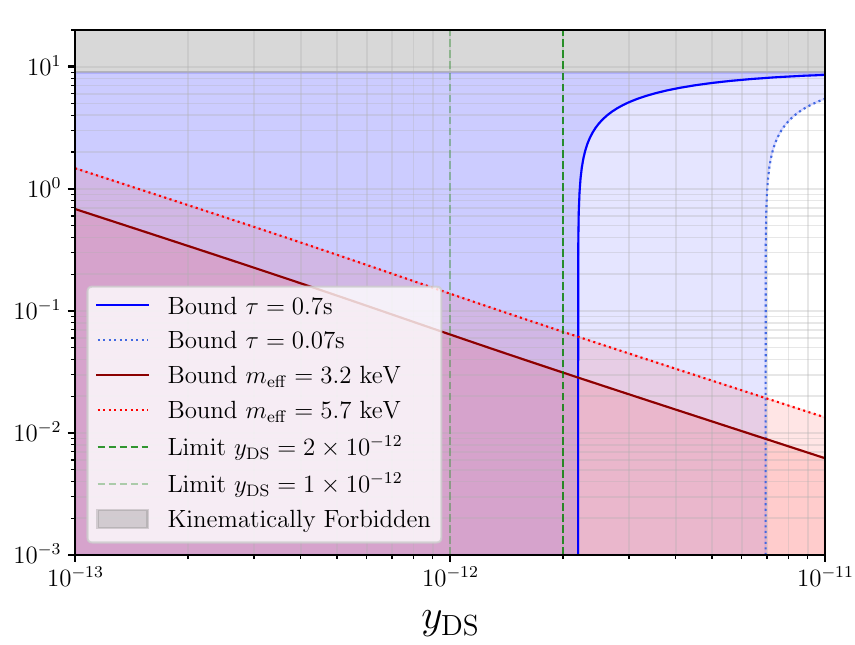} 
        \caption{$m_\phi=10\,\mathrm{GeV}$}
    \end{subfigure}
    
    \caption{Constraints on the $(y_{\mathrm{DS}}, m_\chi)$ parameter space for fixed values of $m_\phi$. The upper panels show the cases where $m_\phi = 1000\,\mathrm{GeV}$ with $m_N = 1, 100$, and $500\,\mathrm{GeV}$, while the lower panels correspond to $m_\phi = 500, 100$, and $10\,\mathrm{GeV}$ with a fixed $m_N = 1\,\mathrm{GeV}$. The line styles and color conventions are identical to those in Fig.~\ref{fig:results2}.}
    \label{fig:results3}
\end{figure}

Next, we fix $m_\phi$ to examine the constraints on the $y_{\mathrm{DS}}$ and $m_\chi$ parameter space, as depicted in the upper panel in the Fig.~\ref{fig:results3}. 
Since the phase space of $\phi$ decay is dominated by $m_\phi$ when $m_\phi \gg m_\chi$ and $m_\phi \gg m_N$,
the BBN limits appear as  nearly straight vertical lines in most of these plots. 
For the velocity dispersion constraints, similarly, an increase in $y_{\mathrm{DS}}$ implies that the DM undergoes more redshifting at later times, 
which in turn allows for a smaller $m_\chi$. Thus, $y_{\mathrm{DS}}$ and $m_\chi$ are inversely correlated in the lines of this constraint.

In the lower panel of Fig.~\ref{fig:results3}, we examine the parameter space for $m_\phi = 500, 100, 10\,\mathrm{GeV}$ with a fixed $m_N = 1\,\mathrm{GeV}$. 
We find that as $m_\phi$ decreases, the BBN bounds shift towards larger values on the $y_{\mathrm{DS}}$ axis and the allowed parameter space for the DM mass $m_\chi$ shrinks accordingly. 
Moreover, by examining the intersection of the Lyman-$\alpha$ limits and the maximum valid coupling for decay production ($y_{\mathrm{DS}} = 2 \times 10^{-12}$), we can extract the minimum allowed DM mass. As shown in Figs.~\ref{fig:results3}(d)-(f) for $m_\phi = 500, 100$, and $10\,\mathrm{GeV}$, these intersections occur at $m_\chi \approx 0.4\,\mathrm{GeV}, 0.1\,\mathrm{GeV}$, and $0.07\,\mathrm{GeV}$, respectively. These results imply that for the decay production mechanism, the DM mass must satisfy a robust lower bound of $\mathcal{O}(10^{-1})\,\mathrm{GeV}$. This conclusion is consistent with our findings in Fig.~\ref{fig:results1}.

\begin{figure}[tbp]
    \centering
    \begin{subfigure}[b]{0.32\textwidth}
        \centering
        \includegraphics[height=3.8cm]{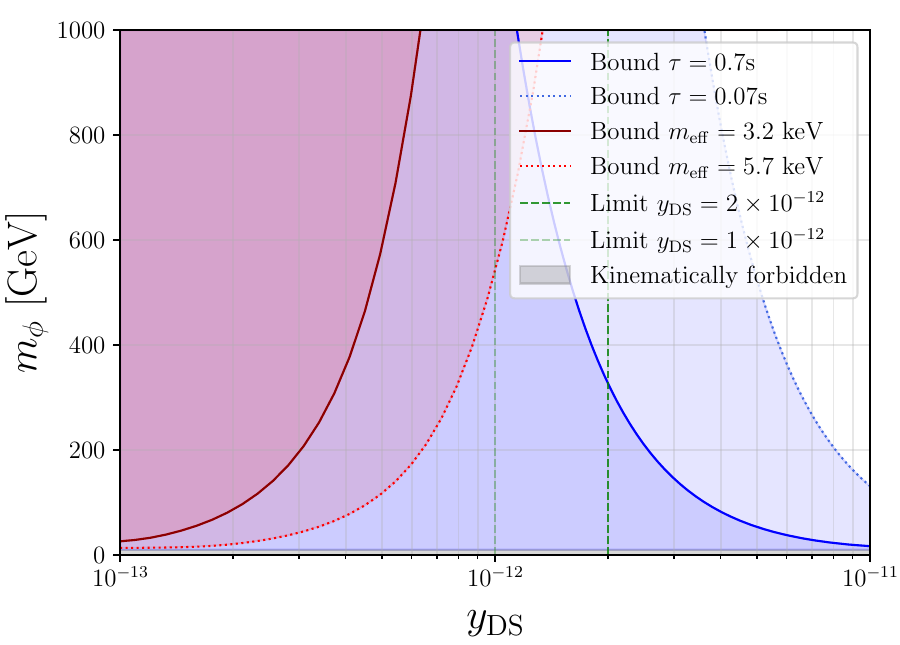} 
        \caption{$m_{N}/m_{\phi}=0.9$}
    \end{subfigure}
    \hfill 
    \begin{subfigure}[b]{0.30\textwidth}
        \centering
        \includegraphics[height=3.8cm]{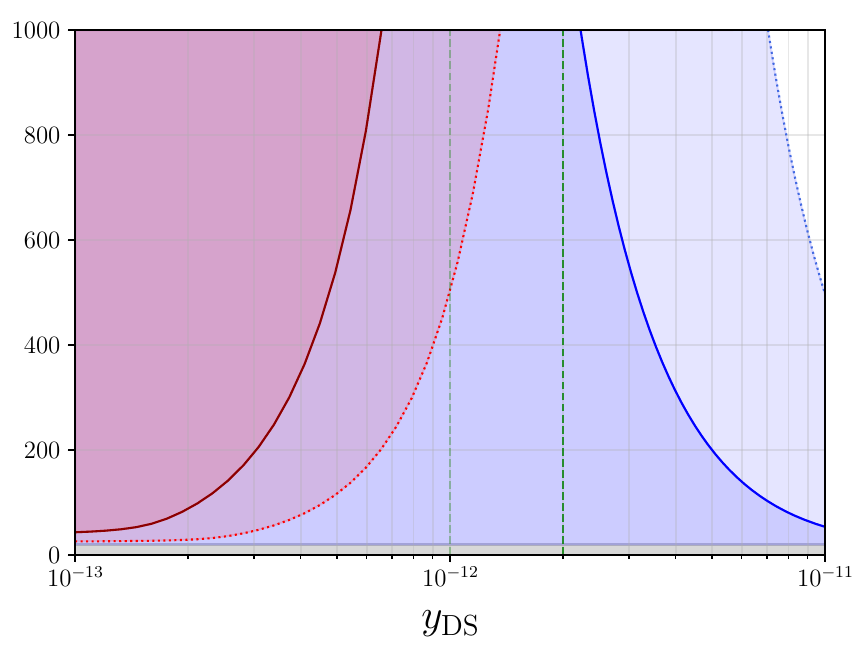} 
        \caption{$m_{N}/m_{\phi}=0.95$}
    \end{subfigure}
    \hfill    
     \begin{subfigure}[b]{0.32\textwidth}
        \centering
        \includegraphics[height=3.8cm]{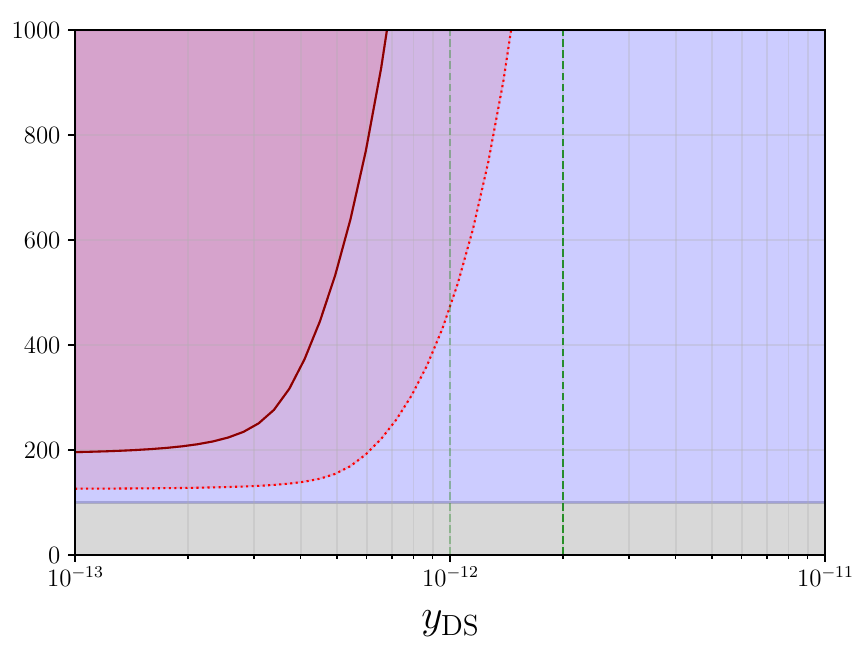} 
        \caption{$m_{N}/m_{\phi}=0.99$}
    \end{subfigure}
    \hfill
    \begin{subfigure}[b]{0.32\textwidth}
        \centering
        \includegraphics[height=3.8cm]{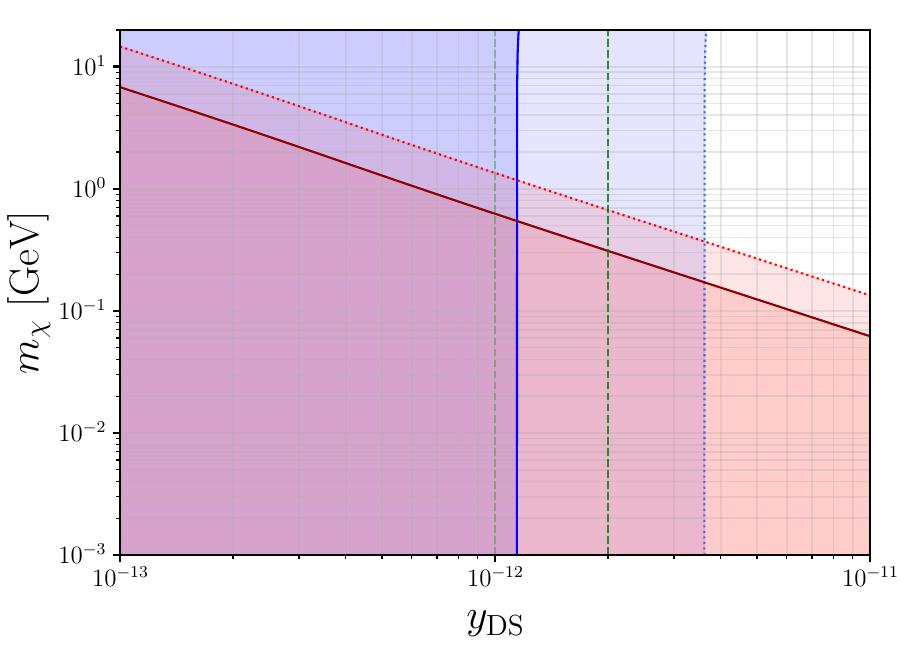} 
        \caption{$m_N/m_\phi=0.9$}
    \end{subfigure}
    \hfill 
    \begin{subfigure}[b]{0.30\textwidth}
        \centering
        \includegraphics[height=3.8cm]{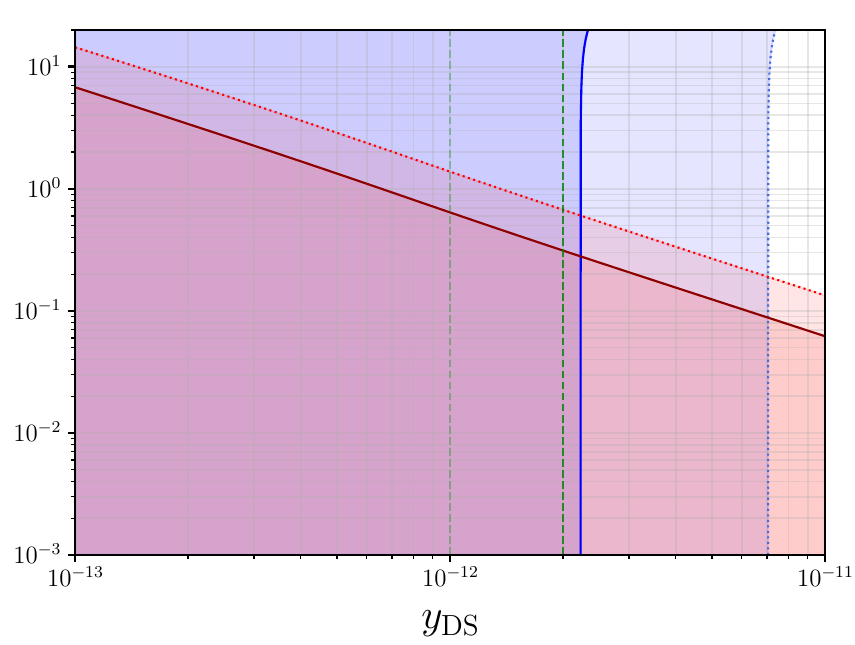} 
        \caption{$m_N/m_\phi=0.95$}
    \end{subfigure}
    \hfill    
     \begin{subfigure}[b]{0.32\textwidth}
        \centering
        \includegraphics[height=3.8cm]{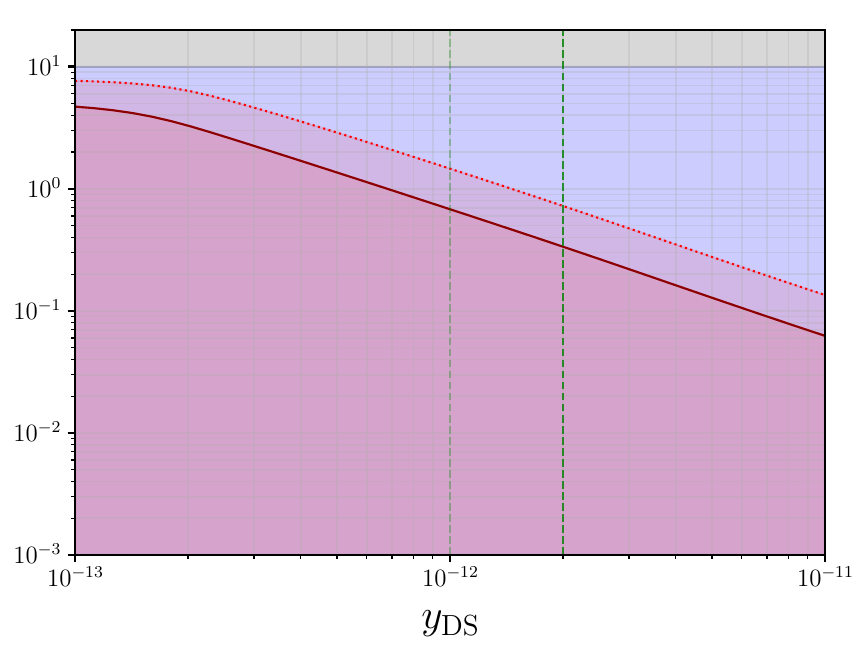} 
        \caption{$m_N/m_\phi=0.99$}
    \end{subfigure}

    \caption{Constraints on the parameter space in the highly degenerate kinematic limit ($m_N/m_\phi \to 1$). 
    Upper panels (a-c): The $(y_{\mathrm{DS}}, m_\phi)$ plane with a fixed DM mass $m_\chi = 1\,\mathrm{GeV}$. 
    Lower panels (d-f): The $(y_{\mathrm{DS}}, m_\chi)$ plane with a fixed parent particle mass $m_\phi = 1000\,\mathrm{GeV}$. In both rows, the mass ratios $m_N/m_\phi$ are set to $0.9$, $0.95$, and $0.99$ from left to right. 
    The line styles and color conventions are identical to those in Fig.~\ref{fig:results2}.}
    \label{fig:Bound_large_mN}
\end{figure}

Finally, we explore the extreme kinematic limit where the masses of the dark scalar and the sterile neutrino are highly degenerate, i.e., $m_N/m_\phi \to 1$.
As first illustrated in upper panels of Fig.~\ref{fig:Bound_large_mN}, we examine the $(y_{\mathrm{DS}}, m_\phi)$ plane by fixing $m_\chi = 1\,\mathrm{GeV}$ and varying $m_N/m_\phi \in \{0.9, 0.95, 0.99\}$. Increasing $m_N/m_\phi$ induces two competing effects on the final DM velocity: a suppressed decay rate $\Gamma_\phi$ which reduces redshifting and increases velocity, and a smaller initial momentum which decreases velocity at later times. 
These two compensating effects leave the velocity dispersion bound(red lines) changes mildly, as can be seen in the upper panel of Fig.~\ref{fig:Bound_large_mN}. Similar effects can also be found in the lower panel of  Fig.~\ref{fig:Bound_large_mN} which examines 
the $(y_{\mathrm{DS}}, m_\chi)$ plane by fixing $m_\phi = 1000\,\mathrm{GeV}$ and varying $m_N/m_\phi \in \{0.9, 0.95, 0.99\}$. 

On the contrary, the BBN constraints change significantly when varying $m_N/m_\phi$, as shown in the Fig.~\ref{fig:Bound_large_mN}. 
This is because a larger $m_N/m_\phi$ suppresses the phase space of $\phi$ decay and leads to a smaller decay rate.
To keep the decay rate large enough and the lifetime of $\phi$ short enough to evade the BBN bounds, large coupling $y_{\mathrm{DS}}$ is required.
Consequently, the BBN constraints tighten monotonically with increasing $m_N/m_\phi$. 

\begin{figure}[htbp]
    \centering
    \captionsetup[subfigure]{justification=centering}
    \begin{subfigure}[b]{0.32\textwidth}
        \centering
        \includegraphics[height=3.8cm]{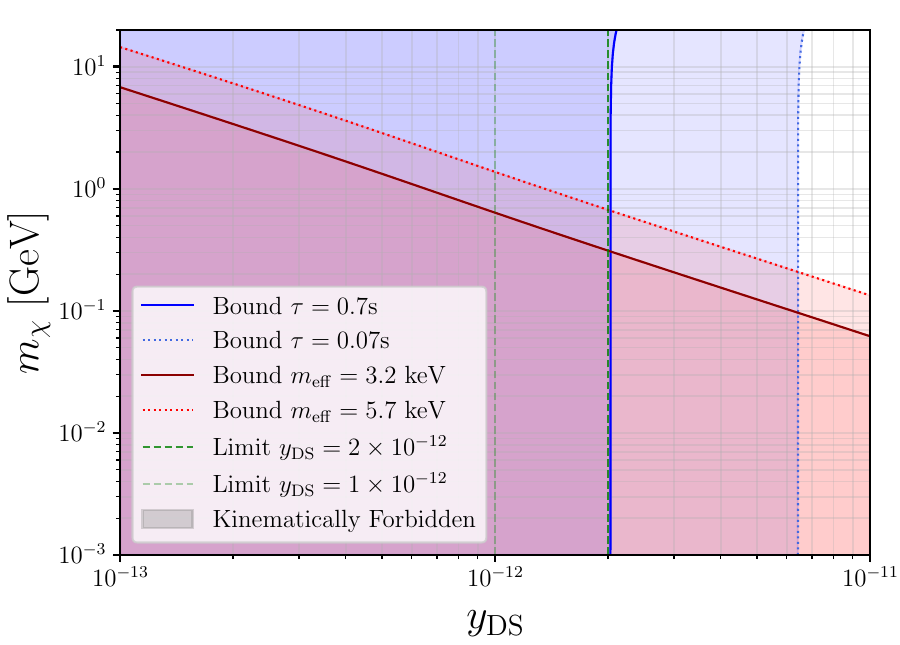} 
        \caption{$m_\phi=1000\,\mathrm{GeV}$, \\ $m_N/m_\phi=0.945$}
    \end{subfigure}
    \hfill 
    \begin{subfigure}[b]{0.30\textwidth}
        \centering
        \includegraphics[height=3.8cm]{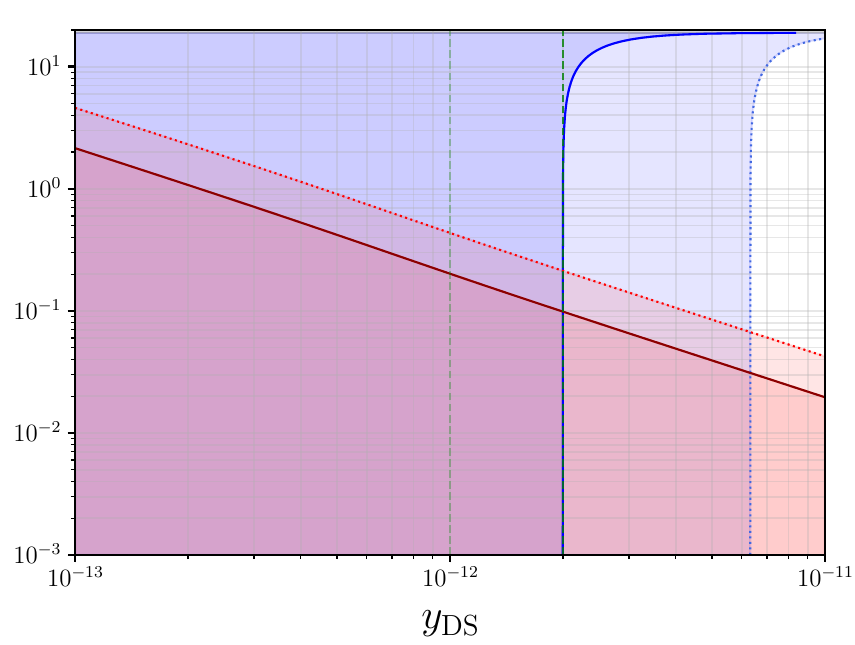} 
        \caption{$m_\phi=100\,\mathrm{GeV}$, \\ $m_N/m_\phi=0.810$}
    \end{subfigure}
    \hfill    
     \begin{subfigure}[b]{0.32\textwidth}
        \centering
        \includegraphics[height=3.8cm]{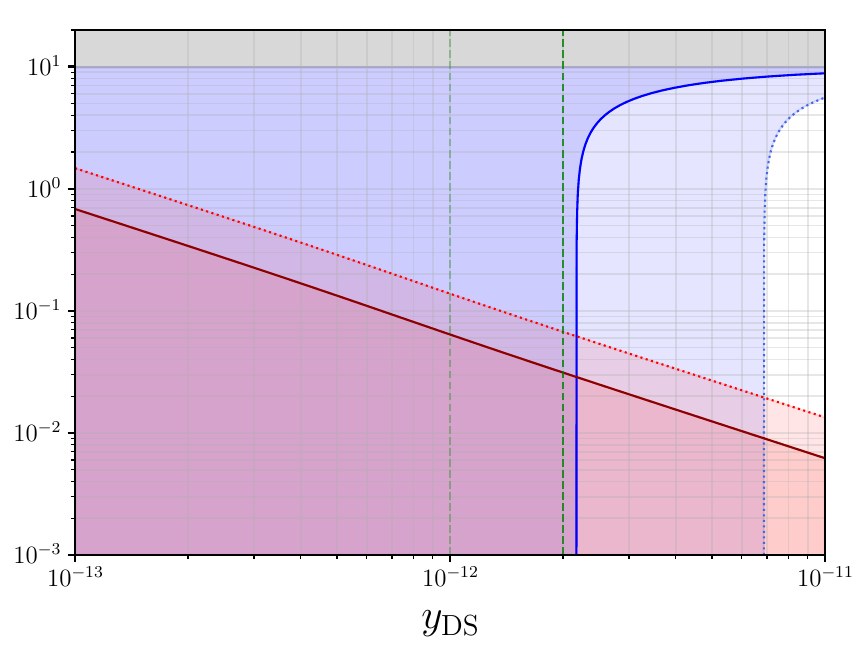} 
        \caption{$m_\phi=10\,\mathrm{GeV}$, \\ $m_N/m_\phi=0.001$}
    \end{subfigure}
    
    \caption{Critical $m_N/m_\phi$ thresholds in the $(y_{\mathrm{DS}}, m_\chi)$ plane where BBN limits completely exclude the decay production. From left to right, the panels correspond to $m_\phi = 1000, 100$, and $10\,\mathrm{GeV}$, with $m_N/m_\phi = 0.945, 0.810$, and $0.001$, respectively. The line styles and color conventions are identical to those in Fig.~\ref{fig:results2}.}
    \label{fig:Bound_rate}
\end{figure}

As illustrated in Fig.~\ref{fig:Bound_rate}, we identify the $m_N/m_\phi$ thresholds where BBN limits completely exclude the decay production scenario for various $m_\phi$. Specifically, for $m_\phi = 1000\,\mathrm{GeV}$, this complete exclusion occurs at $m_N/m_\phi \gtrsim 0.945$. As $m_\phi$ decreases, this threshold drops, occurring at $m_N/m_\phi \gtrsim 0.810$ and $0.001$ for $m_\phi = 100$ and $10\,\mathrm{GeV}$, respectively.
We can see in these plots that the scenario with degenerate masses, i.e. $m_N/m_\phi \sim 1$, is basically excluded by the BBN constraint.

\subsection{Verification via Linear Power Spectrum}
\label{subsec:power_spectrum}

So far, we have obtained the Lyman-$\alpha$ constraint on the light DM from dark sector decay
by mapping the parameter space of our light DM model to an equivalent WDM approximation using the matching of velocity dispersion.
As mentioned in the end of the last section, there is another way of mapping the parameter space of our light DM model to the equivalent WDM approximation,
independently of the method of matching velocity dispersion. 
In this subsection, we are going to explore this transfer function method 
and show that the constraints obtained using these two methods are consistent.

To systematically extract constraints via the transfer function approach, we implement the following step-by-step procedure for a given set of model parameters ($m_\phi, m_\chi, m_N, y_{\mathrm{DS}}$):
\begin{enumerate}
    \item Evaluate the normalized non-thermal distribution $F(q)$ by numerically solving the coupled Boltzmann equations. We then convert $F(q)$ into the standard normalized comoving phase-space distribution $f_\chi(q)$. This derived distribution is directly fed into the Boltzmann solver \texttt{CLASS} to solve the cosmological perturbations and compute the linear matter power spectrum $P_{\mathrm{\chi}}(k)$.
    \item Compute the transfer function $T_{\mathrm{\chi}}(k) = \left[ P_{\mathrm{\chi}}(k) / P_{\mathrm{CDM}}(k) \right]^{1/2}$. We then fit this transfer function using the parameterization in Eq.~(\ref{eq:transfer_func}) with a fixed exponent $\mu = 1.12$ to extract the breaking scale parameter $\alpha_{\mathrm{\chi}}$.
    \item 
    Take $\alpha_{\mathrm{WDM}}=\alpha_{\chi}$ and insert this value into Eq.~(\ref{eq:alpha_WDM}). 
  Map the extracted breaking scale $\alpha_{\mathrm{\chi}}$ to a unique effective thermal mass $m_{\mathrm{eff}}$ by inverting Eq.~(\ref{eq:alpha_WDM}). If $m_{\mathrm{eff}} < m_{\mathrm{WDM}}^{\mathrm{limit}}$ (e.g., $m_{\mathrm{WDM}}^{\mathrm{limit}}=5.7\,\mathrm{keV}$ for the stringent Lyman-$\alpha$ bound), the DM free-streaming effect is overly strong. This leads to an excessive suppression of small-scale structures in the power spectrum, and such parameter points are therefore excluded.
\end{enumerate}

\begin{figure}[htbp]
    \centering
    \begin{subfigure}[b]{0.32\textwidth}
        \centering
        \includegraphics[height=3.8cm]{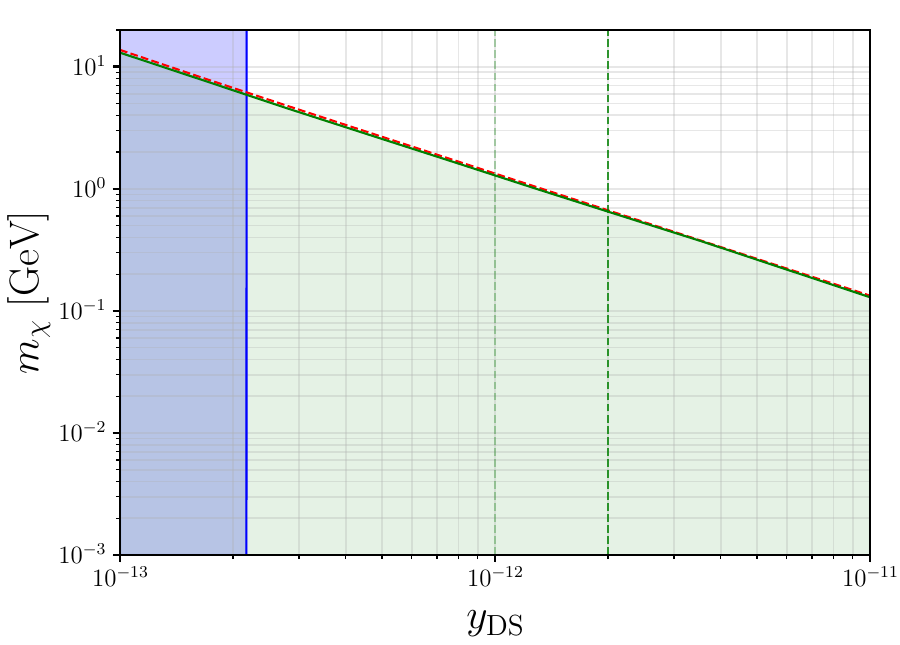} 
        \caption{$m_\phi=10^3\,\mathrm{GeV}$}
    \end{subfigure}
    \hfill 
    \begin{subfigure}[b]{0.30\textwidth}
        \centering
        \includegraphics[height=3.8cm]{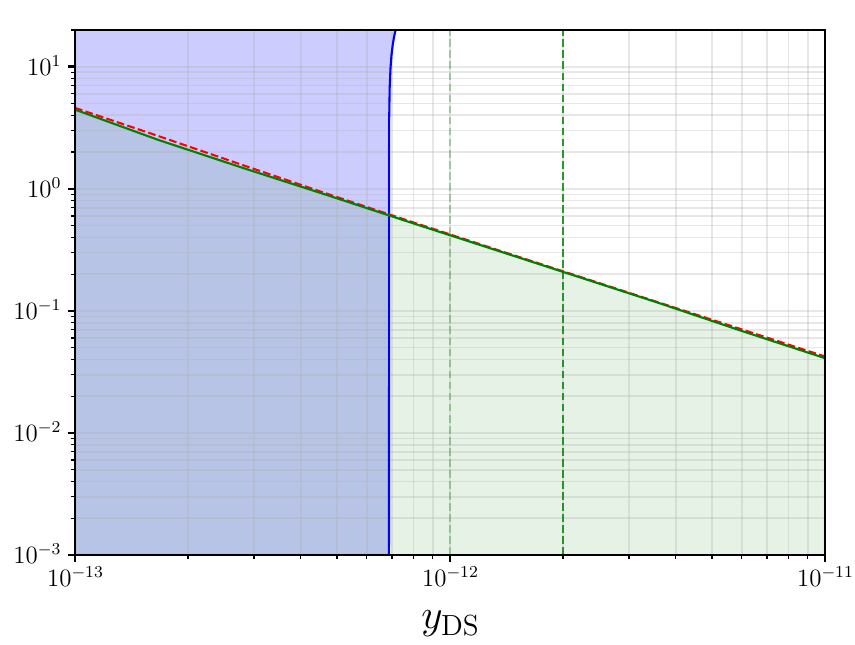} 
        \caption{$m_\phi=100\,\mathrm{GeV}$}
    \end{subfigure}
    \hfill    
     \begin{subfigure}[b]{0.32\textwidth}
        \centering
        \includegraphics[height=3.8cm]{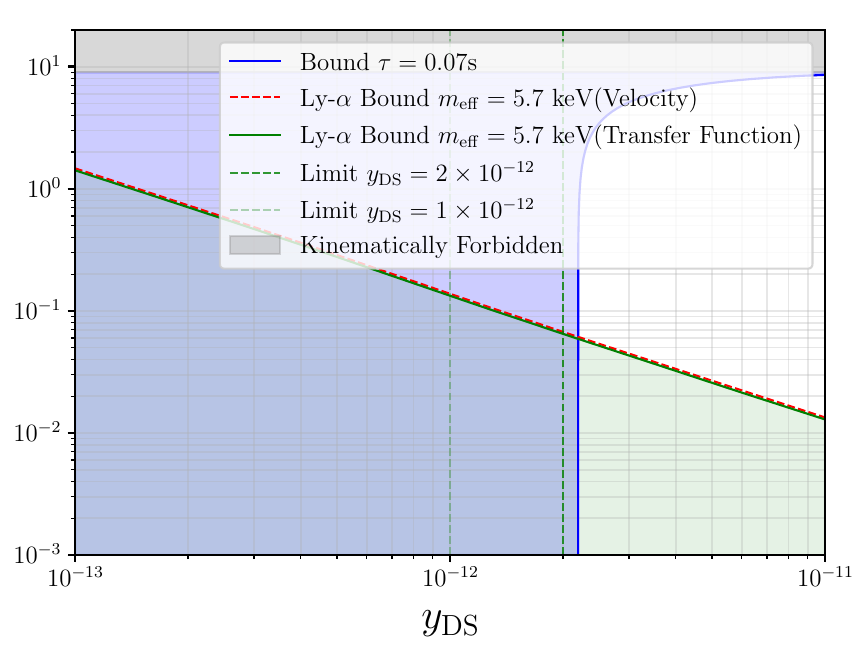} 
        \caption{$m_\phi=10\,\mathrm{GeV}$}
    \end{subfigure}
    
    \caption{Constraints on the parameter space for selected mediator masses $m_\phi = 1000, 100, 10\,\mathrm{GeV}$ with a fixed $m_N = 1\,\mathrm{GeV}$, derived from the transfer function fitting method. The red dashed curve represents the boundary obtained from the velocity dispersion method, while the green solid curve indicates the one derived via the transfer function method.}
    \label{fig:verify}
\end{figure}

Fig.~\ref{fig:verify} illustrates the resulting boundaries on the $(y_{\mathrm{DS}}, m_\chi)$ parameter space. Adopting the same parameter selection as in Fig.~\ref{fig:results3}, we fix $m_N = 1\,\mathrm{GeV}$ and evaluate for selected mediator masses of $m_\phi = 1000, 100$ and $10\,\mathrm{GeV}$. The physical behavior of these limits entirely mirrors our previous analysis. The transfer function constraints dictate an inverse correlation between $y_{\mathrm{DS}}$ and $m_\chi$. A larger $y_{\mathrm{DS}}$ leads to earlier decay and enhanced cosmic redshifting, which suppresses the small-scale power spectrum and thereby accommodates a smaller DM mass. Moreover, as $m_\phi$ decreases from $1000\,\mathrm{GeV}$ to $10\,\mathrm{GeV}$, the allowed parameter space shrinks accordingly. At $m_\phi = 10\,\mathrm{GeV}$, the BBN limits severely compress the viable region.

Ultimately, these two independent methodologies are consistent. In Fig.~\ref{fig:verify}, the boundaries derived via the transfer function method (green solid curves) and the velocity dispersion method (red dashed curves) nearly overlap. As shown in section~\ref{subsec:transfer}, the relative error between the two approaches is bounded within approximately $2\%$ across our parameter space of interest.

\section{Conclusions}
\label{sec:conclusions}

This work presents a comprehensive investigation into the constraining power of Lyman-$\alpha$ forest observations on a specific DM model, where the DM particle $\chi$ is produced from the late-time decays of a frozen-out heavy parent particle $\phi$. By jointly applying the latest Lyman-$\alpha$ limits and BBN bounds, we establish stringent and highly complementary constraints on this DM model. The primary physical conclusions are as follows:

\begin{itemize}
    \item \textbf{Exclusion of sub-GeV DM:} Light DM from late-time decay possesses excessively large velocities. The resulting free-streaming severely washes out small-scale structures. Using the latest Lyman-$\alpha$ bounds, we exclude DM masses $m_\chi \lesssim 0.08\,\mathrm{GeV}$ for the decay production scenario. Furthermore, even the most conservative limits completely rule out the region $m_\chi \lesssim 0.01\,\mathrm{GeV}$. This is the main result of our paper.

    \item \textbf{Tighter lower bounds on the decay coupling $\boldsymbol{y_{\mathrm{DS}}}$:} To suppress the late-time DM velocity, the injected particles must undergo more cosmic redshifting, requiring the parent particle to decay earlier. Combined with BBN bounds, this kinematic requirement pushes the lower bound of $y_{\mathrm{DS}}$ to much higher values, reaching $\mathcal{O}(10^{-12})$. In the sub-GeV DM regime, this joint constraint is significantly stronger than that from BBN alone.

    \item \textbf{Strict upper limits on the parent mass $\boldsymbol{m_\phi}$:} The Lyman-$\alpha$ constraints require the injected DM to remain sufficiently ``cold'', which intrinsically limits the maximum initial decay momentum. Consequently, this imposes an upper bound on the parent particle mass $m_\phi$. In certain mass regimes, this kinematic ceiling tightly squeezes the viable parameter space.
\end{itemize}

In summary, this work highlights the indispensable role of small-scale structure observations in testing non-thermal DM paradigms. By providing stringent constraints that are fundamentally complementary to BBN limits, these observations offer vital guidance for theoretical model building and significantly narrow down the viable parameter space for the DM $\chi$ searches.

\acknowledgments
W. Liao is supported by National Natural Science Foundation of China under the grant No.11875130.

\appendix

\section{Calculation}
\label{app:calculation}

\subsection{Decay Amplitude of $\phi \to N\chi$}
\label{app:calculation.1}

The interaction Lagrangian for the DM model introduced in section~\ref{sec:model} is
\begin{equation}
    \mathcal{L}_{\mathrm{int}} = -y_{\mathrm{DS}} \phi \bar{\chi} N_R + \mathrm{h.c.} \,.
\end{equation}
The spin-averaged squared amplitude for this decay process is calculated as
\begin{equation}
    |\mathcal{M}|^2 = y_{\mathrm{DS}}^2 \sum_{s_\chi, s_N} \bar{u}_\chi P_R u_N \bar{u}_N P_L u_\chi = y_{\mathrm{DS}}^2 \cdot \mathrm{Tr}\left[ (\slashed{p}_N + m_N)P_L(\slashed{p}_\chi + m_\chi)P_R \right] \,.
\end{equation}
Yielding
\begin{equation}
    |\mathcal{M}|^2 = y_{\mathrm{DS}}^2 (m_\phi^2 - m_N^2 - m_\chi^2) \,.
\end{equation}

Next, we evaluate the decay rate by integrating over the phase space
\begin{align}
    \Gamma_\phi &= \frac{1}{2m_\phi} |\mathcal{M}|^2 \int \frac{d^3 p_N}{(2\pi)^3 2E_N} \frac{d^3 p_\chi}{(2\pi)^3 2E_\chi} (2\pi)^4 \delta^{(4)}(p_\phi - p_N - p_\chi) \nonumber \\
    &= \frac{1}{2m_\phi} |\mathcal{M}|^2 \frac{1}{4\pi^2} \int \frac{d^3 p_N}{2E_N 2E_\chi} \delta(m_\phi - E_N - E_\chi) \,.
\end{align}
Evaluating the integral and utilizing the properties of the Dirac delta function, we obtain
\begin{equation}
    \Gamma_\phi = \frac{1}{2m_\phi} y_{\mathrm{DS}}^2 (m_\phi^2 - m_N^2 - m_\chi^2) \frac{|\mathbf{p}_{\mathrm{CM}}|}{4\pi m_\phi} \,,
\end{equation}
where the squared momentum of the decay products in the center-of-mass (CM) frame is 
$|\mathbf{p}_{\mathrm{CM}}|^2 = \frac{1}{4m_\phi^2} \lambda(m_\phi^2, m_N^2, m_\chi^2)$,
with $\lambda(x, y, z) \equiv x^2 + y^2 + z^2 - 2xy - 2xz - 2yz$ being the Källén function. Substituting this back, we arrive at the rest-frame decay rate
\begin{equation}
    \Gamma_\phi = \frac{y_{\mathrm{DS}}^2 m_\phi}{16\pi} \left[ 1 - \left(\frac{m_\chi}{m_\phi}\right)^2 - \left(\frac{m_N}{m_\phi}\right)^2 \right] \lambda^{1/2}\left(1, \frac{m_\chi^2}{m_\phi^2}, \frac{m_N^2}{m_\phi^2}\right) \,.
\end{equation}

\subsection{Derivation of the Number Density Evolution of $\phi$}
\label{app:calculation.2}

At late times, the evolution of the parent particle $\phi$ is dominated by its two-body decay. corresponding collision term is given by
\begin{align}
    C_{\phi \to N+\chi} &= -\frac{1}{2E_\phi} \int \frac{d^3 p'}{(2\pi)^3 2E_N} \int \frac{d^3 q'}{(2\pi)^3 2E_\chi} |\mathcal{M}|^2 (2\pi)^4 \delta^{(4)}(p - p' - q') f_\phi(p) \nonumber \\
    &= - \frac{m_\phi}{E_\phi} \Gamma_\phi f_\phi(p) \,.
    \label{eq:app_C_phi}
\end{align}
Integrating the collision term over the phase space yields the Boltzmann equation for the number density $n_\phi$:
\begin{equation}
    \dot{n}_\phi + 3H n_\phi \approx - \Gamma_\phi \int \frac{d^3 p}{(2\pi)^3} \frac{m_\phi}{E_\phi} f_\phi(p) \,.
    \label{eq:app_ndot_exact}
\end{equation}
As discussed in Appendix~\ref{app:approximation.1}, $\phi$ is highly non-relativistic during this decay epoch ($T \ll m_\phi$, $E_\phi \approx m_\phi$). Thus, the phase space integral simply reduces to the number density $n_\phi$
\begin{equation}
    \dot{n}_\phi + 3H n_\phi \approx - \Gamma_\phi n_\phi \,.
    \label{eq:app_ndot_approx}
\end{equation}
In the comoving framework, the physical number density is related to the comoving number density by $n_\phi = a^{-3} N_\phi$. Therefore, Eq.~\eqref{eq:app_ndot_approx} becomes
\begin{equation}
    \dot{N}_\phi \approx - \Gamma_\phi N_\phi \,.
    \label{eq:app_Ndot_approx}
\end{equation}
Solving this differential equation yields a simple exponential solution
\begin{equation}
N_\phi(t) = N_{\phi, 0} e^{-\Gamma_\phi t}\,,
\label{eq:app_N_solution}
\end{equation}
where the initial amplitude constant $N_{\phi, 0}$ is determined by Eq.~\eqref{eq:Omega}.

\subsection{Derivation of the Collision Term of $\chi$}
\label{app:calculation.3}

For the DM particle $\chi$, the collision term is given by
\begin{align}
    C_\chi[f(p)] &= \frac{1}{2E_\chi(p)} \int \frac{d^3 p'}{(2\pi)^3 2E_N(p')} \int \frac{d^3 q'}{(2\pi)^3 2E_\phi(q')} |\mathcal{M}|^2 (2\pi)^4 \delta^{(4)}(p - p' - q') f_\phi(p) \nonumber \\
    &= \frac{1}{32\pi^2 E_\chi} \int \frac{d^3 p_\phi}{E_\phi E_N} \delta(E_\phi - E_\chi - E_N) |\mathcal{M}|^2 f_\phi(p_\phi) \,.
\end{align}
The angular integration over $\theta$ (the angle between $\mathbf{p}_\phi$ and $\mathbf{p}_\chi$) yields
\begin{equation}
    I_\theta = \int_{-1}^1 d\cos\theta \frac{1}{E_N} \delta\left( E_N(p_\phi, p_\chi, \cos\theta) - (E_\phi - E_\chi) \right) = \frac{1}{p_\phi p_\chi} \,.
\end{equation}
The kinematic limits $\cos\theta \in [-1, 1]$ impose integration bounds on the momentum
\begin{equation}
    \left| E_\chi - E_\phi \frac{E_\chi^*}{m_\phi} \right| \leq p_\phi \frac{p^*}{m_\phi} \,,
\end{equation}
which can be translated to bounds on the comoving momentum $q_\phi$
\begin{equation}
    \left| m_\phi E_\chi - E_\phi E_\chi^* \right| \leq q_\phi \frac{p^*}{a(t)} \,,
\end{equation}
where $p^*$ is the decay momentum of $\chi$ in the rest frame of $\phi$. Therefore, the collision term simplifies to
\begin{equation}
    C_\chi[f(p)] = \frac{1}{32\pi^2 E_\chi} \int \frac{p_\phi^2 dp_\phi}{E_\phi} |\mathcal{M}|^2 f_\phi(p_\phi) \frac{2\pi}{p_\phi p_\chi} = \frac{m_\phi^2 \Gamma_\phi}{2 p^* E_\chi p_\chi} \int dp_\phi f_\phi(p_\phi) \frac{p_\phi}{E_\phi} \,.
\end{equation}
In the comoving framework, defining $F_\chi(q) = \frac{q^3}{2\pi^2} f_\chi(q)$, we finally obtain
\begin{equation}
    C_\chi[F(q)] = \frac{q_\chi^3}{2\pi^2} \times \frac{1}{a} \frac{m_\phi^2 \Gamma_\phi}{2 p^* E_\chi q_\chi} \int dq_\phi f_\phi(q_\phi) \frac{q_\phi}{E_\phi} = \frac{m_\phi^2 \Gamma_\phi q_\chi^2}{2 a p^* E_\chi} \int d\ln q_\phi \frac{F_\phi(q_\phi)}{E_\phi q_\phi} \,.\label{eq:C_chi_F}
\end{equation}
Substituting these derived collision terms back into the general Boltzmann equation, we obtain Eq.~(\ref{eq:Cchi_F}).

To verify the consistency of the derived collision terms, we can check the comoving particle number conservation for the decay process $\phi \to N + \chi$. For $\phi$ particle that decays, one $\chi$ particle is produced. Therefore, the integrated collision terms over the comoving phase space must satisfy
\begin{equation}
    \int d\ln q_\phi \, C_\phi[F(q_\phi)] + \int d\ln q_\chi \, C_\chi[F(q_\chi)] = 0 \,.
\end{equation}
Using the collision term for $\chi$ given in Eq.~\eqref{eq:C_chi_F}, its integral over $d\ln q_\chi$ is
\begin{equation}
    \int d\ln q_\chi \, C_\chi[F(q_\chi)] = \frac{m_\phi^2 \Gamma_\phi}{2 a p^*} \int d\ln q_\phi \frac{F_\phi(q_\phi)}{E_\phi q_\phi} \int dq_\chi \frac{q_\chi}{E_\chi} \,,
\end{equation}
where we have swapped the order of integration. The inner integral over $q_\chi$ can be evaluated by changing the integration variable to the physical energy $E_\chi$, using the relation $q_\chi dq_\chi = a^2 p_\chi dp_\chi = a^2 E_\chi dE_\chi$. The physical kinematic bounds for $E_\chi$ at a fixed $p_\phi$ are $E_\chi^{\pm} = (E_\phi E_\chi^* \pm p_\phi p^*) / m_\phi$. Thus, the inner integral yields
\begin{equation}
    \int dq_\chi \frac{q_\chi}{E_\chi} = a^2 \int_{E_\chi^-}^{E_\chi^+} dE_\chi = a^2 (E_\chi^+ - E_\chi^-) = a^2 \left( \frac{2 p_\phi p^*}{m_\phi} \right) = \frac{2 a q_\phi p^*}{m_\phi} \,.
\end{equation}
Therefore the double integral gives
\begin{align}
    \int d\ln q_\chi \, C_\chi[F(q_\chi)] &= \frac{m_\phi^2 \Gamma_\phi}{2 a p^*} \int d\ln q_\phi \frac{F_\phi(q_\phi)}{E_\phi q_\phi} \left( \frac{2 a q_\phi p^*}{m_\phi} \right) \nonumber \\
    &= \int d\ln q_\phi \frac{m_\phi \Gamma_\phi}{E_\phi} F_\phi(q_\phi) \nonumber \\
    &= - \int d\ln q_\phi \, C_\phi[F(q_\phi)] \,.
\end{align}
This explicitly confirms that the comoving number density is conserved in our formulation, i.e., $dN_\chi/dt = - dN_\phi/dt$.

\section{Approximations}
\label{app:approximation}

\subsection{Zero-momentum Approximation}
\label{app:approximation.1}

Because the decay occurs sufficiently late, the residual thermal momentum of $\phi$ is generally negligible. We consider the possible kinetic histories of $\phi$. 

If the elastic scattering term $C_{\phi\,\mathrm{SM} \to \phi\,\mathrm{SM}}$ is inefficient, $\phi$ would kinetically decouple from the thermal bath early. As freely propagating particles, their physical momentum simply redshifts with the cosmic expansion ($p_\phi \propto a^{-1}$).
Conversely, if this scattering remains efficient, as with typical WIMPs, $\phi$ is kept in kinetic equilibrium with the thermal bath long after its freeze-out ($p_\phi \sim \sqrt{m_\phi T} \propto a^{-1/2}$). In this scenario, its phase space distribution maintains a Maxwell-Boltzmann profile, $f_\phi(p_\phi) \propto e^{-(E_\phi - \mu)/T}$. Here, the chemical potential $\mu(T)$ acts solely as a time-dependent amplitude. Specifically, its initial value at the onset of the decay is fixed by the requirement to reproduce the observed DM relic density, as given by Eq.~(\ref{eq:Omega}). Consequently, we simply model $\phi$ with this thermal profile, whose normalization is governed by comoving particle conservation ($N_{\phi} + N_{\chi}$ = constant). 

Comparing the two scenarios, the kinetic-equilibrium assumption clearly corresponds to a case where $\phi$ retains a larger residual momentum. Therefore, by conservatively modeling $\phi$ with this thermal profile, we safely overestimate its thermal momentum, ensuring that the zero-momentum approximation remains robust.

To validate this approximation across the viable parameter space, we evaluate two extreme kinematic scenarios:
(1) the maximum decay rate scenario, which corresponds to the earliest possible decay and thus the hottest $\phi$; 
(2) the scenario with the minimum ratio between the intrinsic decay momentum $p_{\chi,0}$ and the thermal momentum $p_\phi$, which corresponds to the maximum relative perturbation of the DM intrinsic momentum by thermal fluctuations.

\begin{figure}[tbp]
    \centering
        \includegraphics[width=0.5\textwidth]{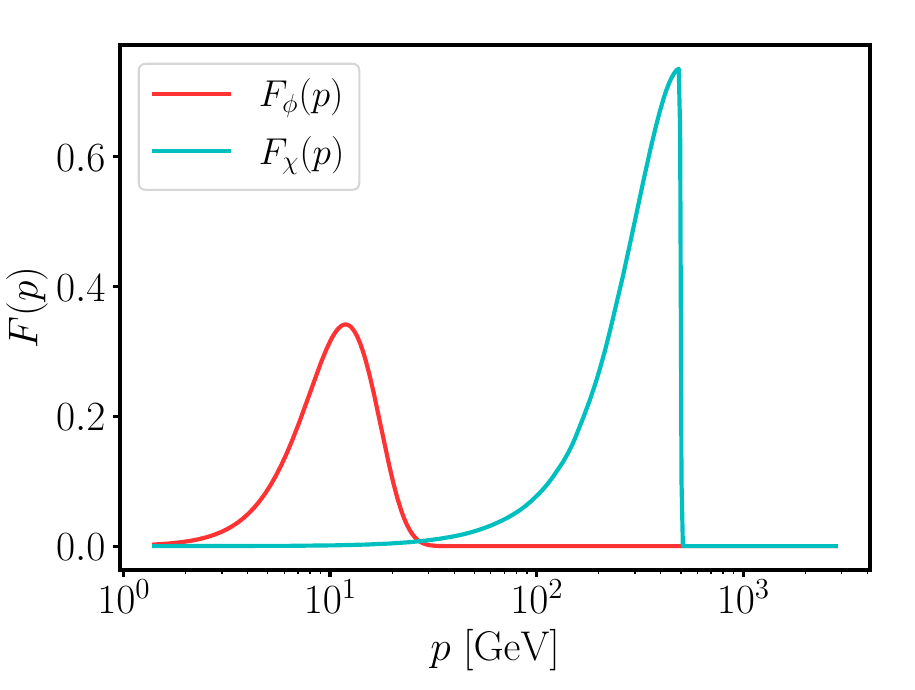} 
    \caption{Phase space distributions of $\phi$ and $\chi$ evolved to the lifetime $t = 1/\Gamma_\phi$. The masses are fixed at $m_\phi = 1000\,\mathrm{GeV}$, $m_N = 1\,\mathrm{GeV}$, $m_\chi = 0.01\,\mathrm{GeV}$, and $y_{\mathrm{DS}} = 10^{-11}$.}
    \label{fig:decay}
\end{figure}

\begin{figure}[htbp]
    \centering
        \includegraphics[width=0.5\textwidth]{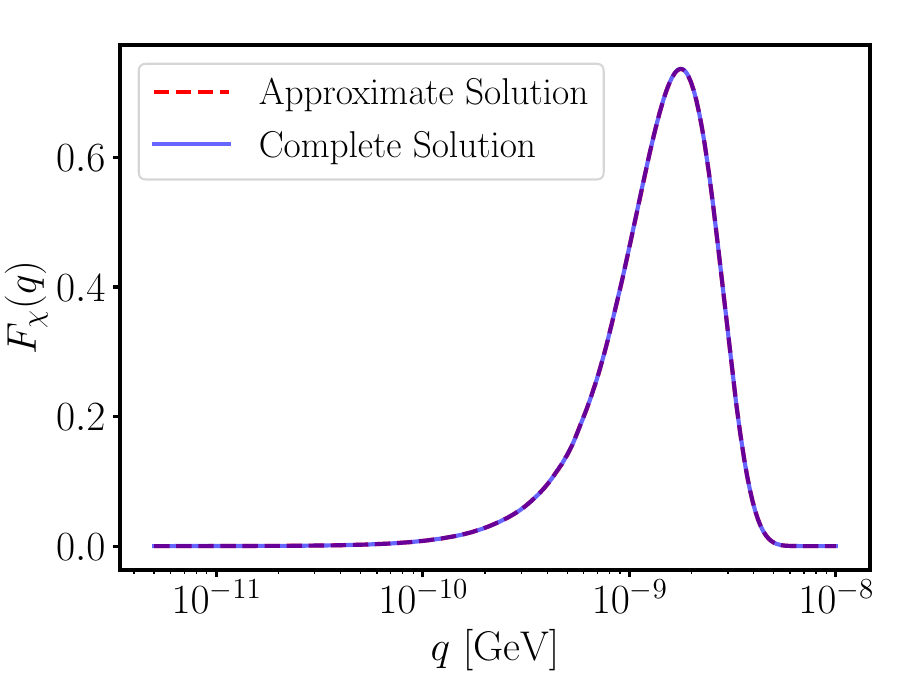} 
    \caption{Comparison of the phase space distribution $F_{\chi}(q)$ obtained from the complete numerical solution and the zero-momentum approximation. The parameter choices are identical to those in Fig.~\ref{fig:decay}.}
    \label{fig:0momentum}
\end{figure}

\paragraph{Maximum decay rate.} 
We select $m_\phi = 1000\,\mathrm{GeV}$, $m_N = 1\,\mathrm{GeV}$, $m_\chi = 0.01\,\mathrm{GeV}$, and $y_{\mathrm{DS}} = 10^{-11}$. This configuration yields the maximum momentum of $\phi$ at decay time, allowing us to safely assess the negligibility of relativistic effects. Fig.~\ref{fig:decay} illustrates the physical momentum distribution of the dark sector particles at decay time. The thermal velocity of $\phi$ is suppressed to $v_\phi \sim \mathcal{O}(10^{-4})$, resulting in a relativistic correction of $\sim \mathcal{O}(10^{-8})$.
Therefore, by adopting the zero-momentum approximation, the collision term simplifies to
\begin{equation}
    C_\chi[F(q)] = \Gamma_\phi N_\phi \cdot \delta(\ln q - \ln q_{\mathrm{inject}}(t)) \,, \label{eq:Cchi_approx}
\end{equation}
where $q_{\mathrm{inject}}(t) = a(t) p^*$ represents the comoving momentum of the $\chi$ particles injected at time $t$. Fig.~\ref{fig:0momentum} demonstrates the phase space distribution $F_{\chi}(q)$ obtained under the zero-momentum approximation and that of the full numerical solution. The difference in $\langle v^2 \rangle$ between the two methods is on the order of numerical noise.


\paragraph{Minimum momentum ratio.} 
To ensure the approximation holds universally, the intrinsic decay momentum $p_{\chi,0}$ must robustly dominate the thermal fluctuation $p_\phi$. Assuming a negligible $N$ mass ($m_N \ll m_\phi$), the intrinsic decay momentum is $p_{\chi,0} = m_\phi(1 - r^2)/2$, where $r \equiv m_\chi/m_\phi$. In the highly degenerate limit ($r \to 1$), the phase-space suppression reduces the decay rate to $\Gamma_\phi = \frac{y_{\mathrm{DS}}^2 m_\phi}{16\pi} (1 - r^2)^2$. 
However, this extreme kinematic limit is vetoed by BBN,

\begin{equation}
    \Gamma_\phi \gtrsim \Gamma_{\mathrm{BBN}} \sim \mathcal{O}(10^{-24})\,\mathrm{GeV}\,.
    \label{eq:Gamma_BBN}
\end{equation}

To maximize the mass degeneracy $r$ while satisfying BBN, we adopt the upper limits of our parameter space: $y_{\mathrm{DS}} = 10^{-11}$ and $m_\phi = 10^3\,\mathrm{GeV}$. Solving the constrained decay rate $\Gamma_\phi \gtrsim \Gamma_{\mathrm{BBN}}$ yields $1 - r^2 \gtrsim \mathcal{O}(10^{-2})$, establishing the upper bound
\begin{equation}
    r_{\mathrm{max}} \sim 0.99.
    \label{eq:rmax_bound}
\end{equation}
At this most extreme scenario, the decay occurs during the BBN epoch at $T \sim 10^{-3}\,\mathrm{GeV}$. Assuming $\phi$ remains in kinetic equilibrium yields its maximum possible thermal momentum. The thermal momentum of $\phi$ is $p_\phi \sim m_\phi \sqrt{T/m_\phi} \sim \mathcal{O}(1)\,\mathrm{GeV}$, while the intrinsic momentum of $\chi$ is $p_{\chi,0} = \frac{m_\phi}{2}(1 - r^2) \sim \mathcal{O}(10^1) \mathrm{GeV}$. Even in this extreme scenario,
\begin{equation}
    \frac{p_{\chi,0}}{p_\phi} \sim \mathcal{O}(10^1)\,.
    \label{eq:momentum_ratio}
\end{equation}
Therefore, the zero-momentum approximation holds well and the mass ratio in our considered scenarios must be restricted to $m_\chi/m_\phi \lesssim 0.99$. Furthermore, as shown in Sec.~\ref{sec:results}, the derived observational bounds lie safely away from this boundary. 

In conclusion, the zero-momentum approximation is firmly justified across the entire viable parameter space.

\subsection{Equivalent WDM Approximation}
\label{app:approximation.2}

To simplify the the constraints from Lyman-$\alpha$ , we map our non-thermal DM model to an equivalent thermal WDM counterpart. As introduced in the section~\ref{sec:imprint}, the effective parameters are determined by matching the background energy density and the comoving pseudo-velocity dispersion, governed by eqs.~(\ref{eq:abundance}) and (\ref{eq:momentum}). 

Requiring the $\chi$ particles to fully account for the observed DM relic abundance, the density matching condition becomes
\begin{equation}
    m_{\chi} \cdot n_{\chi} = \Omega_{\mathrm{DM}} \rho_{c,0} = m_{\mathrm{eff}} \cdot n_{\mathrm{eff}}(T_{\mathrm{eff}}) \,. \label{eq:wdm_relic}
\end{equation}

By utilizing these two matching relations alongside the Fermi-Dirac distribution properties, we have two equations with two degrees of freedom. This allows us to analytically solve for the equivalent temperature $T_{\mathrm{eff}}$ and the effective thermal mass $m_{\mathrm{eff}}$,
\begin{equation}
    T_{\mathrm{eff}} = \left[ \langle v^2 \rangle \frac{(\Omega_{\mathrm{DM}} \rho_{c,0})^2}{C_p C_n^2} \right]^{1/8} \,, \label{eq:T_eff}
\end{equation}
\begin{equation}
    m_{\mathrm{eff}} = \left( \Omega_{\mathrm{DM}} \rho_{c,0} \right)^{1/4} C_n^{-1/4} C_p^{3/8} \langle v^2 \rangle^{-3/8} \,, \label{eq:m_eff}
\end{equation}
where $\langle v^2 \rangle \equiv \langle q^2 \rangle_{\chi} / m_{\chi}^2$ represents the comoving pseudo-velocity dispersion of the DM particles and $C_n$ and $C_p$ are dimensionless constants determined by the Fermi-Dirac statistics, which can be evaluated analytically in terms of the Riemann zeta function $\zeta(x)$,
\begin{align}
    C_n &= \frac{1}{T^3} \int \frac{d^3p}{(2\pi)^3} \frac{2}{\exp(p/T) + 1} = \frac{3\zeta(3)}{2\pi^2} \approx 0.1827 \,, \label{eq:Cn_def} \\
    C_p &= \frac{1}{T^2} \frac{\int \frac{d^3p}{(2\pi)^3} \frac{2p^2}{\exp(p/T) + 1}}{\int \frac{d^3p}{(2\pi)^3} \frac{2}{\exp(p/T) + 1}} = \frac{15\zeta(5)}{\zeta(3)} \approx 12.939 \,. \label{eq:Cp_def}
\end{align}
Substituting these numerical constants back into eqs.~(\ref{eq:T_eff}) and~(\ref{eq:m_eff}), the effective WDM parameters simplify to
\begin{align}
    T_{\mathrm{eff}} &\approx 1.11 \left[ \langle v^2 \rangle (\Omega_{\mathrm{DM}} \rho_{c,0})^2 \right]^{1/8} \,, \\
    m_{\mathrm{eff}} &\approx 3.95 \left( \Omega_{\mathrm{DM}} \rho_{c,0} \right)^{1/4} \langle v^2 \rangle^{-3/8} \,.
\end{align}
These final expressions explicitly demonstrate that both the effective temperature and the effective mass of the equivalent WDM are uniquely determined by the DM velocity dispersion $\langle v^2 \rangle$ and the present-day relic density.

\bibliographystyle{JHEP}  
\bibliography{refs}

@book{Dodelson:2020,
    author    = "Dodelson, Scott and Schmidt, Fabian",
    title     = "{Modern Cosmology}",
    edition   = "2nd",
    publisher = "Academic Press",
    year      = "2020",
    address   = "Amsterdam"
}

@article{FISW:2021,
    author = "Decant, Quentin and Heisig, Jan and Krauss, Martin E. and Ma, Tanguy",
    title = "{Lyman-$\alpha$ constraints on freeze-in and superWIMPs}",
    journal = "JCAP",
    volume = "03",
    pages = "041",
    year = "2022",
    doi = "10.1088/1475-7516/2022/03/041",
    eprint = "2111.09321",
    archivePrefix = "arXiv",
    primaryClass = "hep-ph"
}

@article{Rev:2021,
    author = "Du, Yong and Huang, Fei and Li, Hao-Lin and Li, Yuan-Zhen and Yu, Jiang-Hao",
    title = "{Revisiting Dark Matter Freeze-in and Freeze-out through Phase-Space Distribution}",
    journal = "JCAP",
    doi = "10.1088/1475-7516/2022/04/012",
    volume = "04",
    pages = "012",
    year = "2022",
    eprint = "2111.01267",
    archivePrefix = "arXiv",
    primaryClass = "hep-ph"
}

@article{ChengYu:2021,
    author = "Cheng, Yu and Liao, Wei",
    title = "{Light dark matter from dark sector decay}",
    eprint = "2012.01875",
    archivePrefix = "arXiv",
    primaryClass = "hep-ph",
    journal = "Phys. Lett. B",
    volume = "815",
    pages = "136118",
    year = "2021",
    doi = "10.1016/j.physletb.2021.136118"
}

@article{Planck2018,
    author = "Aghanim, N. and others",
    collaboration = "Planck",
    title = "{Planck 2018 results. VI. Cosmological parameters}",
    eprint = "1807.06209",
    archivePrefix = "arXiv",
    primaryClass = "astro-ph.CO",
    doi = "10.1051/0004-6361/201833910",
    journal = "Astron. Astrophys.",
    volume = "641",
    pages = "A6",
    year = "2020"
}

@article{LZ:2022lsv,
    author = "Aalbers, J. and others",
    collaboration = "LZ",
    title = "{First Dark Matter Search Results from the LUX-ZEPLIN (LZ) Experiment}",
    eprint = "2207.03764",
    archivePrefix = "arXiv",
    primaryClass = "hep-ex",
    doi = "10.1103/PhysRevLett.131.041002",
    journal = "Phys. Rev. Lett.",
    volume = "131",
    number = "4",
    pages = "041002",
    year = "2023"
}

@article{ATLAS:2021kxv,
    author = "Aad, Georges and others",
    collaboration = "ATLAS",
    title = "{Search for new phenomena in events with an energetic jet and missing transverse momentum in $pp$ collisions at $\sqrt{s} = 13$ TeV with the ATLAS detector}",
    eprint = "2102.10874",
    archivePrefix = "arXiv",
    primaryClass = "hep-ex",
    doi = "10.1103/PhysRevD.103.112006",
    journal = "Phys. Rev. D",
    volume = "103",
    number = "11",
    pages = "112006",
    year = "2021"
}

@article{Fermi-LAT:2016uux,
    author = "Albert, A. and others",
    collaboration = "Fermi-LAT, DES",
    title = "{Searching for Dark Matter Annihilation in Recently Discovered Milky Way Satellites with Fermi-LAT}",
    eprint = "1611.03184",
    archivePrefix = "arXiv",
    primaryClass = "astro-ph.HE",
    doi = "10.3847/1538-4357/834/2/110",
    journal = "Astrophys. J.",
    volume = "834",
    number = "2",
    pages = "110",
    year = "2017"
}

@article{Feng:2003xh,
    author = "Feng, Jonathan L. and Rajaraman, Arvind and Takayama, Fumihiro",
    title = "{Superweakly interacting massive particles}",
    eprint = "hep-ph/0302215",
    archivePrefix = "arXiv",
    doi = "10.1103/PhysRevLett.91.011302",
    journal = "Phys. Rev. Lett.",
    volume = "91",
    pages = "011302",
    year = "2003"
}

@article{Irsic2024,
    author = "Ir\v{s}i\v{c}, Vid and others",
    title = "{Unveiling dark matter free-streaming at the smallest scales with the high redshift Lyman-$\alpha$ forest}",
    eprint = "2309.04533",
    archivePrefix = "arXiv",
    primaryClass = "astro-ph.CO",
    doi = "10.1103/PhysRevD.109.043511",
    journal = "Phys. Rev. D",
    volume = "109",
    number = "4",
    pages = "043511",
    year = "2024"
}

@article{Murgia:2017,
    author = "Murgia, R. and Merle, A. and Viel, M. and Totzauer, M. and Schneider, A.",
    title = "{``Non-cold'' dark matter at small scales: a general approach}",
    doi = "10.1088/1475-7516/2017/11/046",
    journal = "JCAP",
    volume = "11",
    pages = "046",
    year = "2017",
    eprint = "1704.07838",
    archivePrefix = "arXiv",
    primaryClass = "astro-ph.CO"
}

@article{Viel:2005,
    author = "Viel, Matteo and Lesgourgues, Julien and Haehnelt, Martin G. and Matarrese, Sabino and Riotto, Antonio",
    title = "{Constraining warm dark matter candidates including sterile neutrinos and light gravitinos with WMAP and the Lyman-alpha forest}",
    journal = "Phys. Rev. D",
    doi = "10.1103/PhysRevD.71.063534",
    volume = "71",
    pages = "063534",
    year = "2005",
    eprint = "astro-ph/0501562",
    archivePrefix = "arXiv"
}

@article{Banik:2021,
    author = "Banik, Nilanjan and Bovy, Jo and Bertone, Gianfranco and Erkal, Denis and de Boer, T. J. L.",
    title = "{Novel constraints on the particle nature of dark matter from stellar streams}",
    journal = "JCAP",
    volume = "10",
    pages = "043",
    year = "2021",
    doi = "10.1088/1475-7516/2021/10/043",
    eprint = "1911.02663",
    archivePrefix = "arXiv",
    primaryClass = "astro-ph.GA"
}

@article{Liu:2024_JWST,
    author = "Liu, Bin and Shan, Huanyuan and Zhang, Jinglan",
    title = "{New galaxy UV luminosity constraints on warm dark matter from JWST}",
    eprint = "2404.13596",
    archivePrefix = "arXiv",
    primaryClass = "astro-ph.CO",
    doi = "10.3847/1538-4357/ad4ed8",
    journal = "Astrophys. J.",
    volume = "968",
    number = "1",
    pages = "79",
    year = "2024"
}

@article{Lesgourgues:2011re,
    author = "Lesgourgues, Julien",
    title = "{The Cosmic Linear Anisotropy Solving System (CLASS) I: Overview}",
    eprint = "1104.2932",
    archivePrefix = "arXiv",
    primaryClass = "astro-ph.IM",
    month = "4",
    year = "2011"
}

@article{Blas:2011rf,
    author = "Blas, Diego and Lesgourgues, Julien and Tram, Thomas",
    title = "{The Cosmic Linear Anisotropy Solving System (CLASS) II: Approximation schemes}",
    eprint = "1104.2933",
    archivePrefix = "arXiv",
    primaryClass = "astro-ph.CO",
    reportNumber = "CERN-PH-TH-2011-082, LAPTH-010-11",
    doi = "10.1088/1475-7516/2011/07/034",
    journal = "JCAP",
    volume = "07",
    pages = "034",
    year = "2011"
}

@article{XENON:2023cxc,
    author = "Aprile, E. and others",
    collaboration = "XENON",
    title = "{First Dark Matter Search with Nuclear Recoils from the XENONnT Experiment}",
    eprint = "2303.14729",
    archivePrefix = "arXiv",
    primaryClass = "hep-ex",
    doi = "10.1103/PhysRevLett.131.041003",
    journal = "Phys. Rev. Lett.",
    volume = "131",
    number = "4",
    pages = "041003",
    year = "2023"
}

@article{Cheng_CPC:2022,
    author = "Cheng, Yu and Liao, Wei and Yan, Qi-Shu",
    title = "{Collider search of light dark matter model with dark sector decay}",
    eprint = "2109.07385",
    archivePrefix = "arXiv",
    primaryClass = "hep-ph",
    doi = "10.1088/1674-1137/ac538c",
    journal = "Chin. Phys. C",
    volume = "46",
    number = "6",
    pages = "063103",
    year = "2022"
}

@article{Meng:2021kio,
    author = "Meng, Yue and others",
    collaboration = "PandaX-4T",
    title = "{Dark Matter Search Results from the PandaX-4T Commissioning Run}",
    eprint = "2107.13438",
    archivePrefix = "arXiv",
    primaryClass = "hep-ex",
    journal = "Phys. Rev. Lett.",
    volume = "127",
    number = "26",
    pages = "261802",
    year = "2021",
    doi = "10.1103/PhysRevLett.127.261802"
}

@article{Liu:2023_Shining,
    author = "Liu, Ang and Shao, Feng-Lan and Han, Zhi-Long and Jin, Yi and Li, Honglei",
    title = "{Constraints on sterile neutrino portal dark matter from cosmology and collider}",
    eprint = "2212.10043",
    archivePrefix = "arXiv",
    primaryClass = "hep-ph",
    doi = "10.1103/PhysRevD.108.115028",
    journal = "Phys. Rev. D",
    volume = "108",
    number = "11",
    pages = "115028",
    year = "2023",
    note = "[arXiv preprint title: \textit{Shinning Light on Sterile Neutrino Portal Dark Matter from Cosmology and Collider}]"
}

@article{CMBSignature_paper,
    author = "Ghosh, Dilip Kumar and Ghosh, Purusottam and Jeesun, Sk",
    title = "{CMB signature of non-thermal Dark Matter produced from self-interacting dark sector}",
    eprint = "2301.13754",
    archivePrefix = "arXiv",
    primaryClass = "hep-ph",
    doi = "10.1088/1475-7516/2023/07/012",
    journal = "JCAP",
    volume = "07",
    pages = "012",
    year = "2023"
}

@article{nuTHDM_paper,
    author = "Liu, Ang and Shao, Feng-Lan and Han, Zhi-Long and Jin, Yi and Li, Honglei",
    title = "{Sterile neutrino portal dark matter in $\nu$THDM}",
    eprint = "2205.11846",
    archivePrefix = "arXiv",
    primaryClass = "hep-ph",
    doi = "10.1140/epjc/s10052-023-11609-5",
    journal = "Eur. Phys. J. C",
    volume = "83",
    number = "5",
    pages = "415",
    year = "2023"
}

@article{Liu:2025_MW,
    author = "Liu, Jianxiang and Gong, Yan and Liao, Kai",
    title = "{Joint Constraints on Fuzzy and Warm Dark Matter from Satellite Populations of the Milky Way and Andromeda}",
    eprint = "2512.01361",
    archivePrefix = "arXiv",
    primaryClass = "astro-ph.CO",
    journal = "Astrophys. J.",
    year = "2026"
}

@article{Nadler:2024_COZMIC,
    author = "Nadler, Ethan O. and An, Rui and Gluscevic, Vera and Benson, Andrew and Du, Xiaolong",
    title = "{COZMIC. I. Cosmological Zoom-in Simulations with Initial Conditions Beyond Cold Dark Matter}",
    eprint = "2410.03635",
    archivePrefix = "arXiv",
    primaryClass = "astro-ph.CO",
    doi = "10.3847/1538-4357/adceef",
    journal = "Astrophys. J.",
    volume = "986",
    pages = "127",
    year = "2025"
}

@article{Doroshkevich:1984,
    author = "Doroshkevich, A. G. and Khlopov, M. Yu.",
    title = "{Formation of structure in the Universe with unstable neutrinos}",
    journal = "Mon. Not. Roy. Astron. Soc.",
    doi = "10.1093/mnras/211.2.277",
    volume = "211",
    pages = "279--282",
    year = "1984"
}

@article{FI:2025,
    author = {D'Eramo, Francesco and Lenoci, Alessandro and Dekker, Ariane},
    title = {Dark matter freeze-in and small-scale observables: Novel mass bounds and viable particle candidates},
    journal = {Phys. Rev. D},
    volume = {112},
    pages = {116008},
    year = {2025},
    doi = {10.1103/j62q-cvkr},
    eprint = {2506.13864},
    archivePrefix = {arXiv},
    primaryClass = {hep-ph}
}

\end{document}